\documentclass[fleqn,usenatbib]{mnras}

\usepackage{newtxtext,newtxmath}

\usepackage[T1]{fontenc}

\DeclareRobustCommand{\VAN}[3]{#2}
\let\VANthebibliography\thebibliography
\def\thebibliography{\DeclareRobustCommand{\VAN}[3]{##3}\VANthebibliography}

\usepackage{graphicx}	
\usepackage{amsmath}	
\usepackage{hyperref}
\usepackage{derivative}
\usepackage[dvipsnames]{xcolor}

\newcommand{\valpha}{\boldsymbol{\alpha}}
\DeclareMathOperator\Arg{Arg}
\newcommand{\inprept}[1]{#1 (\textit{in prep.})}

\title[GPU Microlensing II]{A GPU Code for Finding Microlensing Critical Curves and Caustics}

\author[Weisenbach]{Luke Weisenbach$^{1,}$\thanks{E-mail: weisluke@alum.mit.edu}
\\
$^{1}$Institute of Cosmology and Gravitation, University of Portsmouth, Dennis Sciama Building, Burnaby Road, Portsmouth, PO1 3FX, UK\\
}

\date{Accepted XXX. Received YYY; in original form ZZZ}

\pubyear{2025}

\begin{document}
\label{firstpage}
\pagerange{\pageref{firstpage}--\pageref{lastpage}}
\maketitle

\begin{abstract}
Advancements in analyses of caustic crossing events in gravitationally microlensed quasars and supernovae can benefit from numerical simulations which locate the caustics in conjunction with the creation of magnification maps. We present a GPU code which efficiently solves this problem; the code is available at \url{https://github.com/weisluke/microlensing/}. We discuss how the locations of the microcaustics can be used to determine the number of caustic crossings and the distances to caustics, both of which can be used to constrain the space of nuisance parameters such as source position and velocity within magnification maps.
\end{abstract}




\section{Introduction}
\label{sec:intro}

Since the discovery of the first gravitationally lensed object by \citet{1979Natur.279..381W}, gravitational lensing has grown into a field ripe with potential for furthering our understanding of distant lensing galaxies \citep{2022arXiv221010790S}, clusters \citep{2024SSRv..220...19N}, and their stellar content, as well as the structure of lensed quasars \citep{2024SSRv..220...14V} and supernovae \citep{2024SSRv..220...13S}. On a macro-scale, gravitational lensing produces an observable number of images of a lensed background object. However, if the source is sufficiently compact, the myriad of stars within a lensing galaxy further split each macroimage into an ensemble of unobservable microimages whose brightnesses change as the source moves, contributing to observed fluctuations in the brightnesses of the macroimages \citep{1981ApJ...244..756Y, 1986ApJ...301..503P}.

Temporal variations in image brightness due to microlensing can be used to constrain the size and structure of the source \citep{2002ApJ...577..615W, 2019MNRAS.486.1944V, 2022A&A...659A..21P, 2024MNRAS.531.1095B}, the mass spectrum of the microlenses \citep{2001MNRAS.320...21W}, and the stellar mass fraction in the lens galaxy \citep{2004ApJ...605...58K, 2009ApJ...693..174C} (or equivalently the dark matter fraction). Occasionally, new microimages can appear due to the relative motion or a change in the size of the source. In these instances, the observed brightness of the source can drastically increase. Such High Magnification Events (HMEs) can be used to probe the structure of the lensed source on nano-arcsecond scales, a possibility which is unique to microlensing. A multitude of gravitationally lensed quasars and supernovae are expected to be discovered in the upcoming decade with the Vera Rubin Observatory (LSST), \textit{Euclid}, and the \textit{Roman} Space Telescope, with hundreds of HMEs occurring per year \citep{2020MNRAS.495..544N}. The wealth of light curve level data from monitoring these systems will allow for unprecedented constraints on quasar and supernovae structure and stellar mass in the lens galaxies, with further constraining power possible on the individual sources assuming early detection and sufficient high resolution followup of HMEs.

The standard tool used in analyzing microlensed light curves is a map of the microlensed magnification as a function of source position \citep{1990PhDT.......180W}. While magnification maps provide direct information about the temporally evolving microlensed magnifications after accounting for the size and movement of the source, they only indirectly provide the precise source locations where new microimages appear and HMEs occur, known as caustics. Incorporating additional information about the caustics can better inform likelihood analyses that substantially improve upon traditional methods by narrowing the allowable space of nuisance parameters such as the source position and trajectory. The importance of combining magnification maps and caustic locations for such purposes was realized already in \citet{1992A&A...258..591W}, though few improvements have appeared in the literature since then until recently. 

Works which have incorporated information about the caustics into supernovae light curve analyses have shown how constraints on the presence or absence of HMEs and the typical timescales of caustic crossings can be used to reduce magnification uncertainties \citep{2024MNRAS.531.4349W} and constrain stellar mass fractions and mean microlens mass \citep{2025arXiv250201728W}. Such constraints from supernovae are similar to those presented in the literature for quasars, albeit with different computational considerations due to the nature of supernovae microlensing (an expanding supernova photosphere versus a moving quasar accretion disk). Both of the above works heavily relied on new numerical methods which were greatly improved through the usage of Graphics Processing Units (GPUs). The purpose of this work is to discuss those numerical methods and make public the code used. Incorporating the methods developed into future analyses of gravitationally microlensed systems will be important for further improving cosmological and astrophysical constraints from lensed supernovae and quasars. In addition, improving the computational techniques available for microlensing analyses is an active area of research in anticipation of the LSST data stream, and it is our hope that adding to the available toolbox might help motivate and stimulate further studies.

In what follows, we provide a short introduction to the phenomena of gravitational (micro)lensing and discuss some aspects of computational microlensing in Section~\ref{sec:theory}. We provide an overview of the methods previously used to determine the locations of microlensing caustics before proceeding to develop and improve the computational techniques in Sections~\ref{sec:finding_ccs} and \ref{sec:comp_improvements}. We discuss how to calculate the number of caustic crossings in Section~\ref{sec:calculating_ncc}, and provide an overview on how the resulting data can be used to determine the distances to caustics for lensed supernovae and quasars. Finally, we present conclusions and discuss the applications of the code to future areas of interest in Section~\ref{sec:conclusions}.

\section{Gravitational (Micro)Lensing Theory and Computation}
\label{sec:theory}

\subsection{Lensing theory}

The phenomenon of gravitational lensing can be succinctly described using one quantity: the time delay of a photon relative to the unlensed case, given by \citep{2024SSRv..220...12S} \begin{equation}
    \tau = \frac{1}{2}\Big(\mathbf{x} - \mathbf{y}\Big)^2 - \psi(\mathbf{x})
\end{equation} where $\mathbf{y}$ is the (unobservable) position of the source on the sky, $\mathbf{x}$ is the position of an image, and $\psi$ is a two dimensional projection of the gravitational potential of the lens. This deceptively simple expression has been non-dimensionalized, removing physical constants and cosmological dependencies which, for our purposes, are unneeded. 

The lens equation, which can be simply written as \begin{equation}
    \nabla \tau = 0
\end{equation} or more fully as \begin{equation}
    \mathbf{y} = \mathbf{x} - \nabla\psi(\mathbf{x}),
\end{equation} provides the locations where the multiple images of the source are seen -- from Fermat's principle, they occur at stationary points of the time delay surface, which can be either minima, maxima, or saddlepoints. The magnifications of the images are inversely proportional to the curvature of the time delay surface at their location and are given by \begin{equation}
    \mu =  1 / \det\left(\frac{\partial\mathbf{y}}{\partial\mathbf{x}}\right),
\end{equation} with the sign of the magnification determining the parity of the image. Saddlepoints have $\mu<0$, while minima and maxima have $\mu>0$. Furthermore, minima are always magnified, whereas saddlepoints and maxima can be demagnified.

In the vicinity of a single image (macroimage), the lens mapping can be simply written as \begin{equation}
    \mathbf{y} = \left(
\begin{array}{cc}1-\kappa+\gamma & 0\\
0 & 1-\kappa-\gamma
\end{array}\right)\mathbf{x}
\label{eq:macro_lens_mapping}
\end{equation} by Taylor expanding the potential and rotating the coordinate system. Here, $\kappa$ denotes the effective convergence of the lens at the position of the macroimage and $\gamma$ denotes its shear. The gravitational potential is related to $\kappa$ through the two-dimensional Poisson equation \begin{equation}
    \kappa(\boldsymbol x) = \frac{1}{2}\nabla^2\psi(\boldsymbol x) = \frac{1}{2}\big(\psi_{11}+\psi_{22}\big),
\end{equation} while $\gamma=\sqrt{\gamma_1^2+\gamma_2^2}$ with \begin{equation}
    \gamma_1 = \frac{1}{2}\big(\psi_{11}-\psi_{22}\big), \quad \gamma_2 = \psi_{12}=\psi_{21}.
\end{equation} Minima of the time delay surface occur when $1 - \kappa - \gamma > 0$, saddlepoints occur when $1 - \kappa - \gamma < 0 < 1 - \kappa + \gamma$, and maxima occur when $1 - \kappa + \gamma < 0$. The magnification of a macroimage can be computed as \begin{equation}\label{eq:macro_mag}
    \mu_{\text{macro}} = \frac{1}{(1-\kappa)^2 - \gamma^2}.
\end{equation} 

A portion of the surface mass density comes from microlenses, which have an effective convergence \begin{equation}
    \kappa_\star = \frac{\pi\theta_\star^2N_\star\langle m \rangle}{A_\star} 
\end{equation} where the $N_\star$ microlenses of mean mass $\langle m_\star \rangle$ (in units of some mass $M$, typically $M_\odot$, that determines a characteristic length $\theta_\star$ known as the Einstein radius) are randomly distributed in some area $A_\star$ around the macroimage position. The lens equation with microlenses included is \begin{equation}
    \mathbf{y} = \left(
\begin{array}{cc}1-\kappa+\gamma & 0\\
0 & 1-\kappa-\gamma
\end{array}\right)\mathbf{x} - \theta_\star^2\sum_{i=1}^{N_\star}\frac{m_i(\bf x - \bf x_i)}{|\bf x - \bf x_i|^2} - \valpha_s(\mathbf{x})
\label{eq:micro_lens_mapping}
\end{equation} where $m_i$ is the mass of an individual microlens and $\bf x_i$ is its position. We must include a compensating (negative) smooth matter deflection term $\valpha_s(\mathbf{x})$ with the same convergence as the microlenses so that the overall macromodel convergence does not change. 

Equation~\eqref{eq:macro_lens_mapping} clearly has one image position $\mathbf{x}$, corresponding to the observed macroimage, for a given source position $\mathbf{y}$. With the introduction of microlenses, the macroimage is broken up and Equation~\eqref{eq:micro_lens_mapping} gains a multitude of additional solutions corresponding to many microimages. Far away from the macroimage location, many solutions are located near microlenses and correspond to faint saddlepoint images. It is only in the vicinity of the macroimage position that bright microminima and microsaddles appear.

As the lens equation is a smooth function and images can have positive or negative parity, the value of $\mu^{-1}=\det\left(\partial\mathbf{y}/\partial\mathbf{x}\right)$ must somewhere change sign from positive to negative. The critical curves of lensing systems are the loci where $\mu$ becomes formally infinite ($\mu^{-1}=0$), while the caustics are the mappings of these loci to the source plane. A source crossing a caustic is accompanied by the creation or annihilation of a pair of images somewhere along the critical curves, inducing a significant change in the brightness of the macroimage (i.e. a HME). The presence of many microlenses creates a vast network of both caustics, which the source moves (quasar) or expands (supernova) through, and critical curves from which microimages are created or upon which they are annihilated. A source located within this microcaustic network can experience changes in brightness of a magnitude or more as it moves or expands through the network.

\subsection{Computational microlensing}

\begin{figure*}
    \centering
    \includegraphics[width=\textwidth]{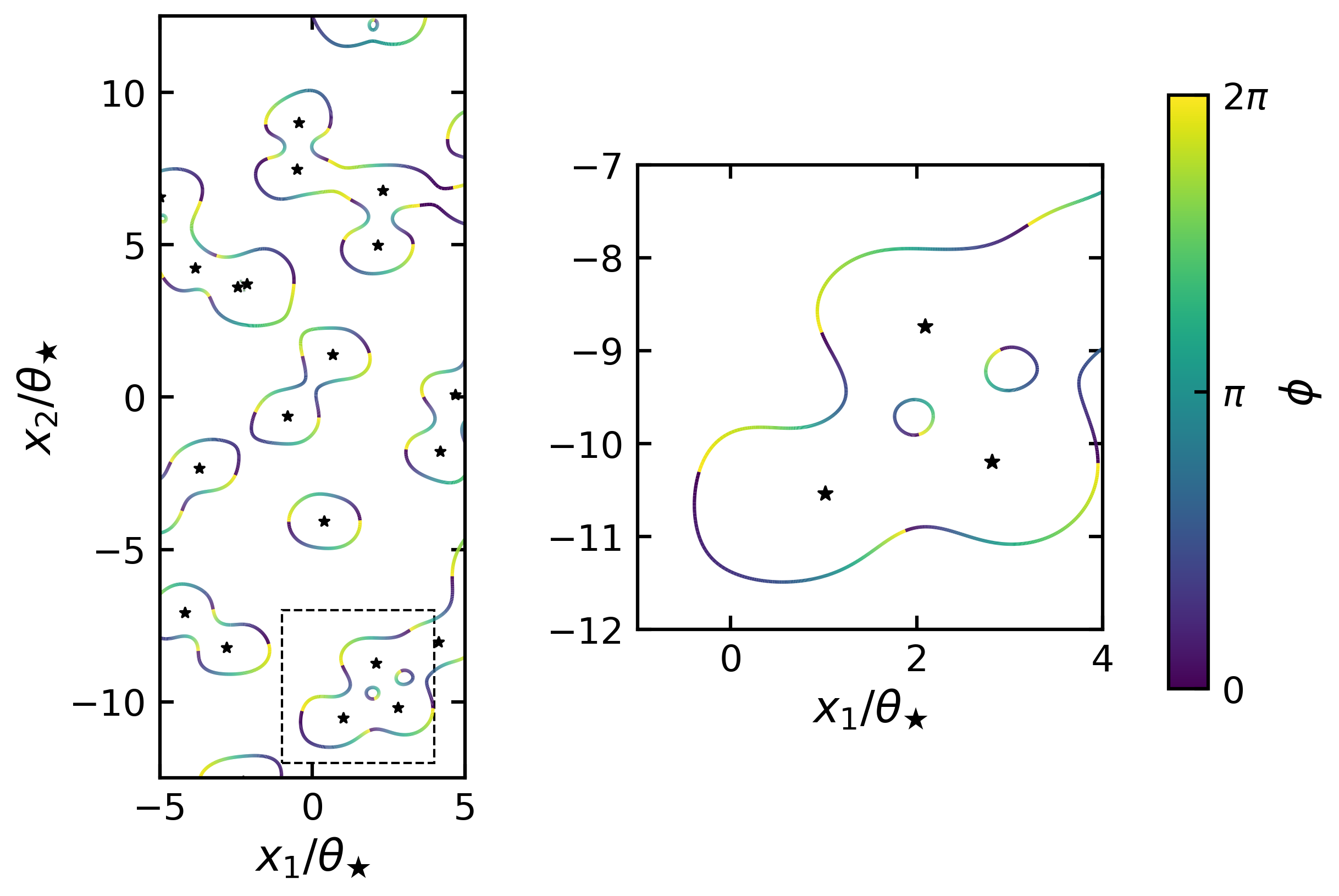}
    \includegraphics[width=\textwidth]{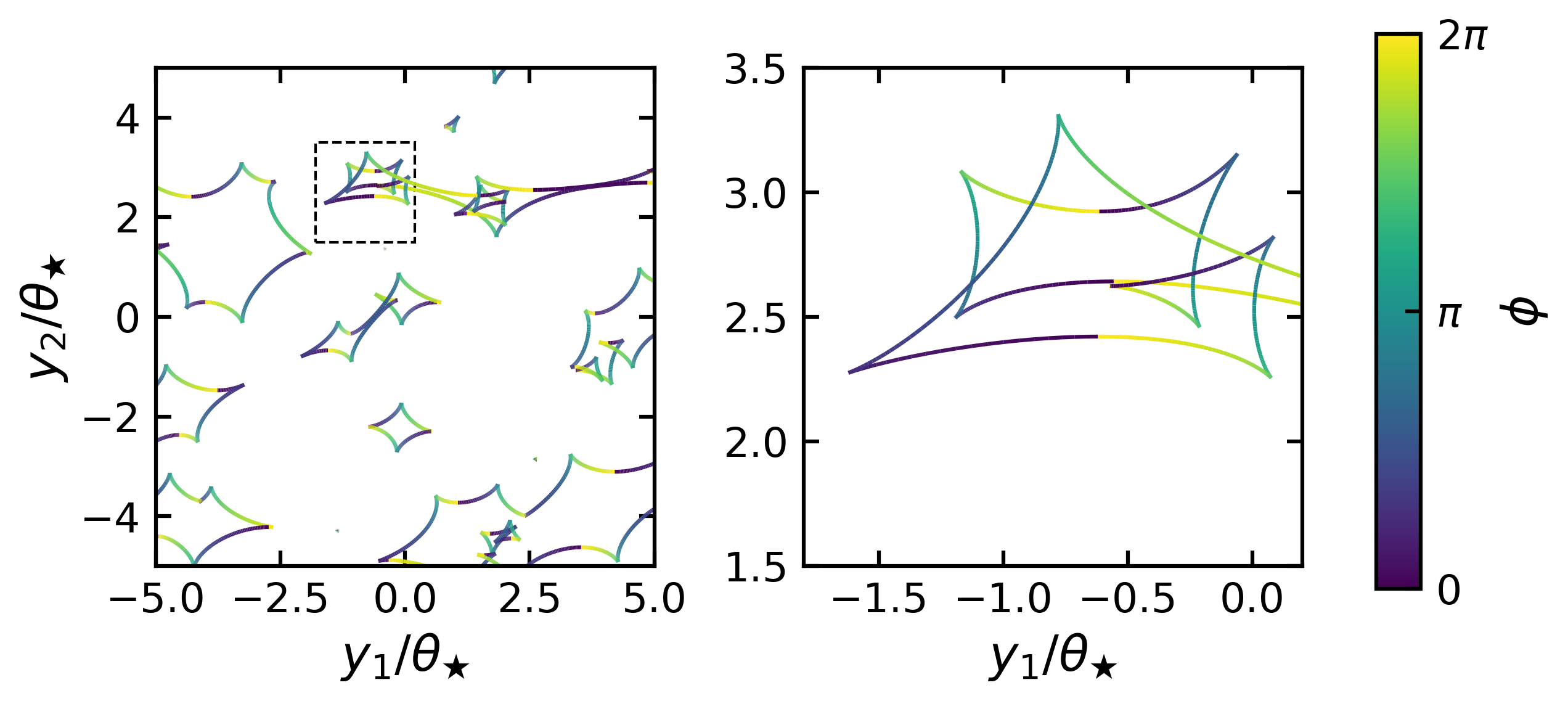}
    \caption{Critical curves (top) and caustics (bottom) for a field of random microlenses. The microlens positions are marked with $\star$ symbols. The rectangular image plane region maps back to the square source plane region; the caustics of the (approximately) single and binary lens critical curves near the origin of the image plane are visible near the origin of the source plane. Points along the critical curves and caustics are color coded according to the value of $\phi$. The right figures are zooms of the dashed squares indicated in the left.}
    \label{fig:crit_curves}
\end{figure*}

Microlensing has long benefited from numerical simulations that allow for examinations of situations where analytic treatment does not suffice. The backbone for many studies are magnification maps that give the total magnification for an object, gravitationally lensed by a field of random microlenses, as a function of the source plane position. The inverse ray shooting (IRS) method for creating such maps was first used in works such as \citet{1986A&A...164..237S} and \citet{1986A&A...166...36K}, and later popularized by \citet{1990PhDT.......180W}. In this method, a multitude of light rays are traced from the image plane to the source plane with the lens equation; the density of light rays in the source plane is proportional to the magnification of the source. In essence, this method sums up at each source pixel the flux contributions of the many microimages of that pixel -- though the ray density must be great enough that no microimages of significant flux are missed. Inverse polygon mapping (IPM) \citep{2006ApJ...653..942M, 2011ApJ...741...42M} provides an alternative to IRS for calculating the magnification maps, with various computational speed ups recently developed \citep{2021A&A...653A.121S, 2022ApJ...941...80J}. 

Magnification maps made with ray shooting by direct calculations of the lensing deflection angle can be time intensive to compute due to the large number of arithmetic operations that must be performed. A vast number of individual light rays must be traced backwards to the source plane in order to achieve good statistics, and each individual microlens deflects a ray by some amount. Tree codes such as the Fast Multipole Method \citep[FMM,][]{1987JCoPh..73..325G} alleviate the number of computations necessary by using a multipole expansion to approximate the gravitational effect of groups of microlenses far away from a given ray location. Some authors have additionally proposed altering the shape of the random microlens field, which is typically taken to be a circle in such simulations, to reduce the number of microlenses, and therefore calculations, required \citep{2022ApJ...931..114Z}. However, the process of shooting many rays is inherently parallelizable -- the deflection angle of one light ray does not depend on another, only on the microlenses and the model macro-parameters. GPUs have been used extensively in scientific applications for parallel processing data, and have allowed for ray tracing simulations to be sped up by many orders of magnitudes \citep{2010NewA...15...16T, 2010NewA...15..726B, teralens}.

Modern consumer GPUs easily operate in the range of hundreds or thousands of GFLOPS for single precision arithmetic. Given that ray tracing only cares about the pixel each light ray lands in, single precision is sufficient and allows for quick computational speed. However, GPUs are also capable of double precision arithmetic operations at slower speeds, while still providing substantial improvements over the amount of time calculations would take with a Central Processing Unit (CPU). GPUs have been used in the context of exoplanet microlensing to speed up various contouring methods for calculating light curves \citep{2011PhDT.......335H}, with recent improvements appearing in \citet{2025ApJS..276...40W} and \citet{2025AJ....169..170R} to recover parameters. However, the extents to which GPUs might be applied to other methods of studying microlensing at the moderate or high optical depth typical for lensed quasars and supernovae have not been greatly explored, to the author's knowledge. Such alternative ways of studying the effects of gravitational microlensing by a random field of microlenses that have received (comparatively) less numerical attention than creating magnification maps include finding individual microimages \citep{1986ApJ...301..503P}, calculating the light curves of gravitationally lensed point sources via the properties of lensed infinite lines \citep{1993ApJ...403..530W, 1993MNRAS.261..647L}, and finding the networks of critical curves and caustics due to a field of microlenses \citep{1990A&A...236..311W}. It is the latter of these to which we devote our attention, as there exist numerical methods which make such calculations very effective via GPU. 

\section{Locating microlensing critical curves and caustics}
\label{sec:finding_ccs}

In this section, we outline the methods by which microlensing critical curves and caustics are located.

\subsection{Parametric Representation of the Critical Curves}

While the lens equation as previously written is a function of positions $\mathbf{x}=(x_1,x_2)$ and $\mathbf{y}=(y_1,y_2)$, it is sometimes simpler to work with complex numbers \citep{1973ApJ...185..747B, 1975ApJ...195...13B, 1990A&A...236..311W} by defining \begin{equation}
    \begin{aligned}
        z &= x_1 + ix_2,\\
        w &= y_1 + iy_2.
    \end{aligned}
\end{equation} The lens equation then becomes a complex function of $z$ and its conjugate $\overline{z}$, and Equation~\eqref{eq:micro_lens_mapping} takes the form \begin{equation}
    w = (1-\kappa)z+\gamma\overline{z}-\theta_\star^{2}\sum_{i=1}^{N_\star}\frac{m_{i}}{\overline{z}-\overline{z_{i}}} - \alpha_s(z).
\end{equation} \citet{1990A&A...236..311W} gives the following parametric representation of the critical curves: \begin{equation}
    \frac{\partial w}{\partial z}e^{i\phi} - \frac{\partial w}{\partial \overline{z}}=0,\hspace{3mm}0\leq\phi<2\pi.
\end{equation} From the lens equation we find the parametric critical curve equation \begin{equation}
    \gamma+\theta_\star^{2}\sum_{i=1}^{N_\star}\frac{m_{i}}{(z-z_i)^2} - \overline{\frac{\partial\alpha_s}{\partial \overline{z}}} - \left(1-\kappa - \frac{\partial\alpha_s}{\partial z}\right)e^{-i\phi}=0
    \label{eq:parametric_cc}
\end{equation} after some conjugation, negation, and using the fact that $\partial\alpha_s / \partial z$ is a real function \citep{1990A&A...236..311W}. For simplicity, we define \begin{equation}
    F(z,\phi) = \gamma+\theta_\star^{2}\sum_{i=1}^{N_\star}\frac{m_{i}}{(z-z_i)^2} - \overline{\frac{\partial\alpha_s}{\partial \overline{z}}} - \left(1-\kappa - \frac{\partial\alpha_s}{\partial z}\right)e^{-i\phi}
\end{equation} and use $F(z)$ to denote the function at an arbitrary value of the phase $\phi$. 

We note that in the context of ray shooting, rays are always shot within the field of microlenses, and so one never has to worry about the behavior of the lens equation or its derivatives at the boundary of the microlens field. This is not the case for the critical curves however, as derivatives of $\alpha_s(z)$ becomes discontinuous at the boundary. Let us assume that these discontinuities can be removed or approximated through various means and that the derivatives are, or can be approximated as, polynomials. We justify this simplification in Appendix~\ref{app:removing_discontinuities}.

We are then interested in the locations where $F(z)=0$. \citet{1990A&A...236..311W} showed that $F(z)$ can be rewritten as a rational function; the numerator is given by \begin{equation}
    P(z)=F(z)\prod_{i=1}^{N_\star}(z-z_i)^2
\end{equation} and its roots are roots of $F(z)$\footnote{For physical reasons, a critical curve cannot pass through the position of a microlens; thus there is no risk of adding spurious roots from this rationalization.}. In general, the roots are all distinct, except possibly at beak to beak or higher order catastrophes\footnote{Whether there actually are any higher order catastrophes or not for a particular value of $\phi$ poses no issues in the following discussions however.}. The number of roots $n_r$ is at least $2N_\star$, plus an additional number that depends on $\alpha_s(z)$.

Each of the roots traces a curve as $\phi$ varies from $0$ to $2\pi$. Once the $n_r$ initial roots are found for $\phi=0$, $\phi$ can be varied from $0$ to $2\pi$ using some number of steps $m$ with $\Delta\phi=\frac{2\pi}{m}$. Roots of $\phi_{j+1}=\phi_j+\Delta\phi$ are in the vicinity of $\phi_j$ for sufficiently small $\Delta\phi$ (sufficiently large $m$) \citep{1990A&A...236..311W}. Figure~\ref{fig:crit_curves} shows the critical curves of a field of microlenses found using $m=200$. 

\subsection{Finding roots}

\begin{figure}
    \centering
    \includegraphics[width=0.5\textwidth]{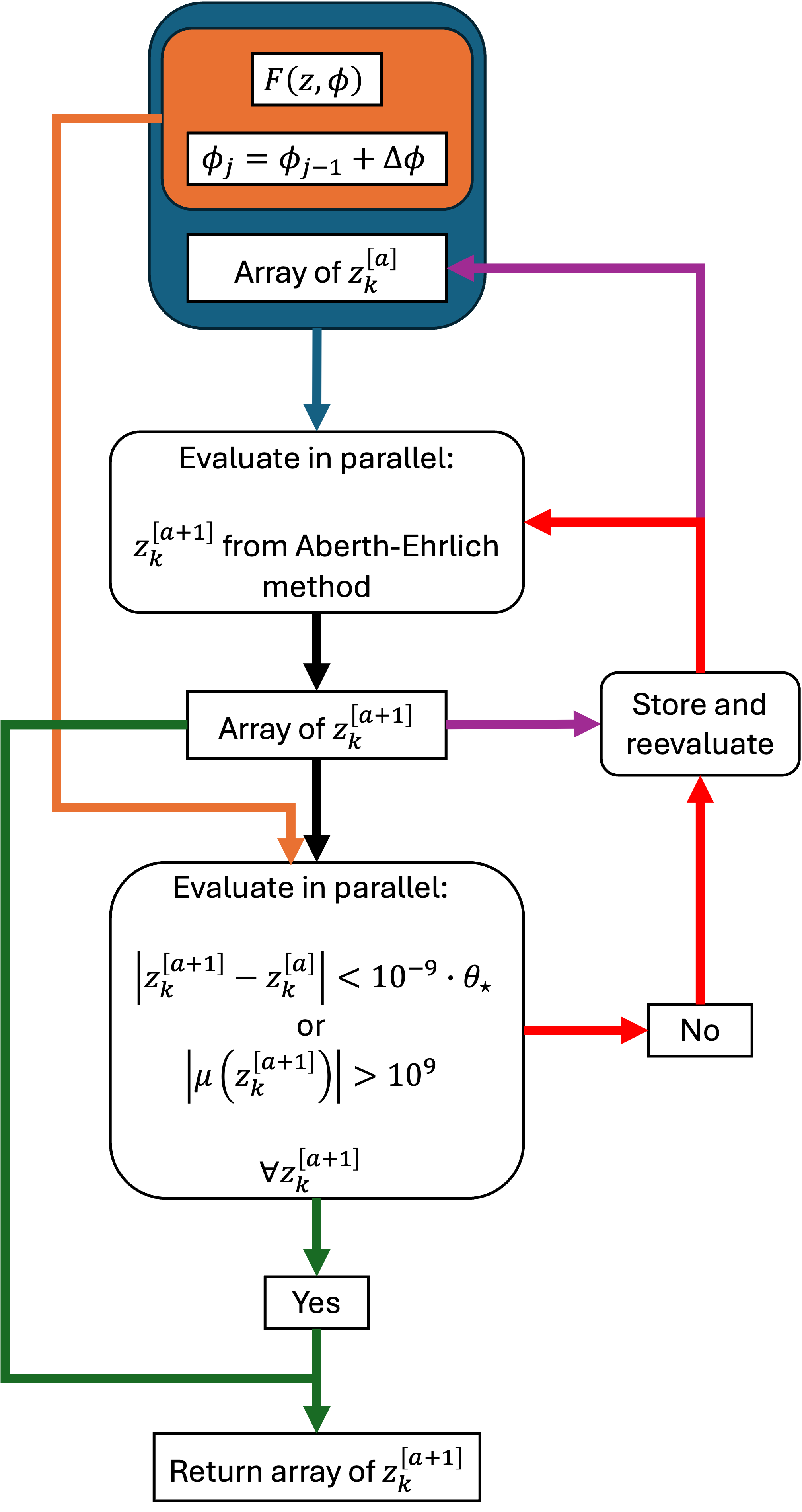}
    \caption{Visualization of the steps needed to find the roots of $F(z,\phi)$ for each value of $\phi$.}
    \label{fig:aberth_ehrlich1}
\end{figure}

The process of first finding the $n_r$ solutions for $\phi=0$, and then following each solution as $\phi$ varies, can be time consuming. This is in part due to the sequential nature of finding the roots in previous works \citep{1990A&A...236..311W, 1993A&A...268..501W}, which largely seem to first find one root to the desired accuracy using Newton's method, divide it out of the polynomial, then find the next root, and so on. The process of finding new roots as $\phi$ changes is sequential as well; one must add $\Delta\phi$ to $\phi_j$, and then sequentially calculate the new $n_r$ roots using the previous roots as starting positions, again using Newton's method or some other iterative root finding scheme. The process can be greatly sped up if one can find the initial $n_r$ roots (and thus, subsequent $n_r$ roots for various values of $\phi$) in parallel. We discuss such a procedure below, which draws wholly from the more general mathematical and computational work of \citet{2017GhiEA}. 

The Aberth-Ehrlich method \citep{1967Ehrlich, 1973Aberth} gives an iterative formula to find an updated particular approximate root $z_k^{[a+1]}$ of a polynomial $p(z)$, using the current particular approximation $z_k^{[a]}$ and all other current approximate roots $z_r^{[a]}$, as \begin{equation}
    z_k^{[a+1]} = z_k^{[a]} - \frac{1}{\frac{p'(z_k^{[a]})}{p(z_k^{[a]})} - \displaystyle{\sum_{r\neq k}}\frac{1}{z_k^{[a]}-z_r^{[a]}} }
\end{equation} where $z_k^{[a]}$ denotes the value of some root $z_k$ at iteration $a$. The first term in the denominator comes from Newton's method, while the second correction term is a sum that depends on the positions of the other roots. To restate the analogy used by others to picture this method \citep{1973Aberth}, the (unknown) exact root positions can be visualized as unmovable positive unit point charges. The approximate root positions are taken as movable negative unit point charges, whose positions are iterated over time. The unknown exact roots act as sinks that draw in the approximate roots. When an approximate root becomes close to an actual root, their charges cancel and there is no longer a draw to that location for any other approximate root. The approximate roots repel each other, preventing any two from converging towards the same position (unless that position is a root of multiplicity greater than one). This iteration scheme is cubically convergent, except where there are multiple roots (for which it converges linearly).

For our microlensing equations, one of the fractions in this method is simplified as follows \citep{1993A&A...268..501W}: \begin{align}
    \frac{P'(z)}{P(z)}&=\frac{d}{dz}\ln P(z) = \frac{d}{dz}\Big(\ln F(z)+2\sum_{i=1}^{N_\star}\ln(z-z_i)\Big)\\
    &=\frac{F'(z)}{F(z)}+2\sum_{i=1}^{N_\star}\frac{1}{z-z_i}
\end{align}

Thus, the Aberth-Ehrlich method in our case will involve some approximate root $z_k$, the values of $F(z)$ and $F'(z)$ at this root, the distances from that root to various microlenses $z_i$, and the distances to various other approximate roots $z_r$.

A parallel implementation of the Aberth-Ehrlich method using NVIDIA's CUDA computing platform is described at length in \citet{2017GhiEA}. We give a brief overview here as it relates to our work.

We start with initial approximations $z_k$ to the $n_r$ roots taken to be the microlens positions $z_i\pm1$, plus an additional number as required from $\alpha_s(z)$ evenly spread in a circle around the microlenses\footnote{Whether or not these are good first approximations is not too important. The main reason we do this is that we know the critical curves lie in or around the field of microlenses, and since our microlenses were already randomly positioned, adding $\pm1$ to their positions should be good enough for our starting guesses.}. Each of these initial roots $z_k^{[0]}$ is stored in an array (of length $n_r$). The advantage to the Aberth-Ehrlich method lies in the fact that for each iteration step $[a]$, the formula can be applied to each approximate root simultaneously, as the only information needed is the function we seek the roots of, its derivative, the microlens positions, and the current set of approximations. Each iteration $[a]$ of a loop calculates the new approximation $z_k^{[a+1]}$ for each root in parallel on the GPU. When a root is sufficiently close to its previous position or gives the desired value of $1/\mu$ close to 0, a flag is set from false to true and that root is no longer iterated\footnote{We use $\big|z_k^{[a+1]}-z_k^{[a]}\big|<10^{-9}\cdot\theta_\star$ or $|\mu|>10^{9}$ as our cutoffs for sufficiently close. The value of $1/\mu$ depends on the value of $F(z,\phi)$; see Appendix~\ref{app:max_mu_error} for an expression giving the maximum possible error independent of $\phi$.}. This process is shown in Figure~\ref{fig:aberth_ehrlich1}. The process stops when all roots are found to the desired accuracy, which we have found takes of order $\sim30-50$ iterations for most number of microlenses $N_\star$ of order $10^3-10^4$. Whereas finding each root to the desired precision one by one through Newton's method might take $10-100$ iterations \textit{per root} depending on the starting guess, the Aberth-Ehrlich method provides all roots with fewer iterations.

Once the $n_r$ roots have been found for $\phi=0$, we must vary $\phi$ from $0$ to $2\pi$. Since the roots for $\phi_j+\Delta\phi$ should be close to those of $\phi_j$, we take the set of solutions for $\phi_j$ as our initial guess in the Aberth-Ehrlich method for the roots of $\phi_{j+1}$. For large $m$, the number of iterations needed goes down from $\sim50$ to $\sim20-30$ for each value of $\phi$.

\section{Computational improvements}
\label{sec:comp_improvements}
In this section, we note a few of the computational improvements which speed up calculations.

\subsection{The fast multipole method}

\citet{1993A&A...268..501W} discuss how computational difficulties can be partially overcome by placing a grid over the field of microlenses and only using microlenses in close proximity to a given grid square (node), with more distant microlenses approximated by some Taylor expansion. This procedure is in essence the FMM of \citet{1987JCoPh..73..325G}, though missing some ideas about the node size and requisite expansion order to maintain accuracy. 

The term \begin{equation}
    \theta_\star^2\sum_{i=1}^{N_\star}\frac{m_i}{(z-z_i)^2} = \frac{\partial^2\psi_\star(z)}{\partial z^2}
\end{equation} depends on all of the microlenses, which can be many for some sets of parameters. We use the FMM to split the potential $\psi_\star$ into two components: one from nearby microlenses $\psi_{\star,\text{ near}}$, and one from distant microlenses $\psi_{\star,\text{ far}}$. The nearby microlenses are used directly, while the distant ones are locally approximated within a node by a Taylor series found from the multipole expansions of distant nodes. The Taylor series for distant microlenses is used in the calculation of $w$, $F(z)$, and $F'(z)$. While the number of calculations for each root from sums which depend on the number of microlenses drastically decreases, we note that we still must find $n_r$ roots in total, which still has a dependency on $N_\star$. 

\subsection{A rectangular microlens field}

If the field of microlenses is circular, $N_\star$ can be very large. The idea of using a rectangular microlens field for ray-shooting simulations in order to reduce the number of microlenses required (of particular interest for highly magnified systems) was first explored in \citet{2022ApJ...931..114Z}. They derive the form of $\valpha_s(\mathbf{x})$ for a rectangular microlens field in vector coordinates. \inprept{Weisenbach} derives the form of $\alpha(z)$ in complex coordinates, from which one can find $\partial\alpha_s(z)/\partial z$ and $\partial\alpha_s(z)/\partial\overline{z}$. Discontinuities in the derivatives of $\alpha_s(z)$ are more prevalent in this case than for a circular field of microlenses, though as Appendices~\ref{app:removing_discontinuities} and \ref{app:a_smooth_approx} show they can be circumvented and approximated with polynomials. There are some computational overheads compared to a circular field of microlenses due to the logarithmic terms present in $\alpha_s(z)$ here, but the decrease in the number of stars reduces $n_r$ and hence can drastically reduce the time required to calculate the positions of the critical curves.

\subsection{Growing the phase chains}

\begin{figure}
    \centering
    \includegraphics[width=0.45\textwidth]{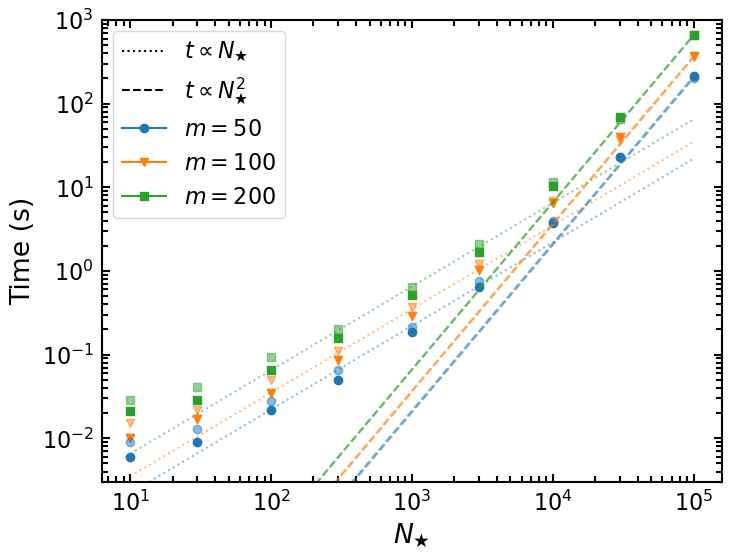}
    \caption{Time taken to calculate critical curves on an NVIDIA A100 80GB GPU, vs. number of microlenses $N_\star$, for various values of the number of steps $m$ used to follow $\phi$ from 0 to $2\pi$. The dashed lines show the asymptotic runtime scaling $t\propto N_\star^2$ of the parallelized Aberth-Ehrlich method, while the dotted lines show the approximate $t\propto N_\star$ of low $N_\star$. The solid markers are for a circular field of microlenses, while the opaque markers are for a rectangular field.}
    \label{fig:time_vs_num_stars}
\end{figure}

\begin{figure*}
    \centering
    \includegraphics[width=\textwidth]{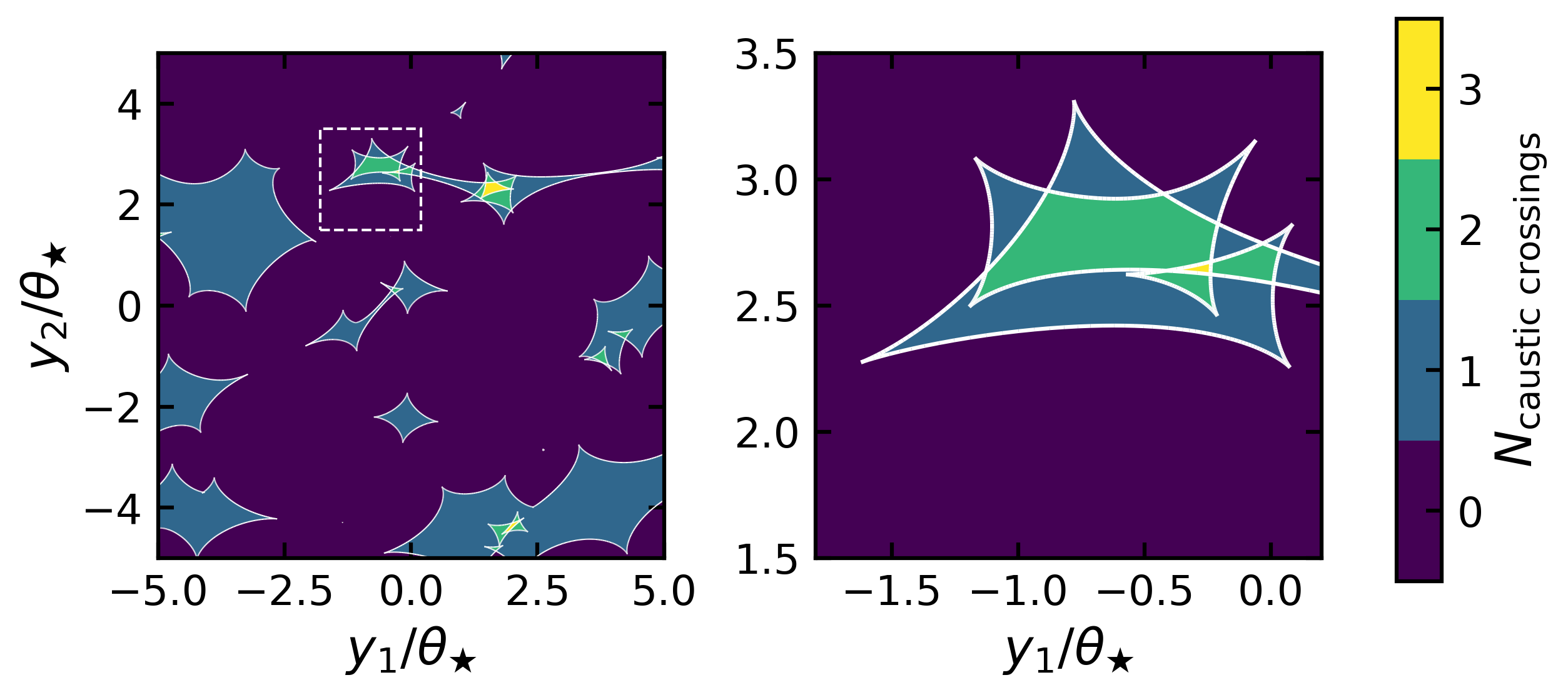}
    \caption{Visualization of the number of caustic crossings. The locations of the caustics are shown with white lines. Pixelation effects near the caustics would be seen with a sufficient zoom, but determining the number of caustic crossings at pixel locations far from the caustics is well defined.}
    \label{fig:ncc}
\end{figure*}

Every one of the initial $n_r$ roots begins and ends a ``chain'' of values from $0$ to $2\pi$. This can be seen in Figure~\ref{fig:crit_curves}, where one can trace a chain (via color, in the figure, via the phase $\phi$ generically) from start to finish. Sometimes a single chain creates a closed critical curve, while sometimes 2 or more chains are needed to fully create a closed critical curve; see the Figure~\ref{fig:crit_curves} zoom. 

The numerical method as described up to this point traces a chain from start to finish sequentially. However, this need not be the case. We further speed up the process by noting that instead of advancing from $0$ to $2\pi$, we may instead grow the chain from the center outwards, going from $\pi$ to $0$ and $2\pi$ independently and in parallel as well, thus reducing the compute time by a factor of $\sim2$ if the GPU threads are not already saturated. For our implementation of the method, we store the $n_r$ initial roots in the central row of a matrix with dimensions $(m+1, n_r)$ where we require $m$ to be even\footnote{We start at $\phi=\pi$ so that a chain covers the range from $0$ to $2\pi$, for simplicity. We use $m+1$ as we wish to include both the starting and ending points of $0$ and $2\pi$ in order to make it easy to determine where the separate chains might join together -- the start of one chain is the end of another, with chains joining together until closed loops are formed.}. For each advancement $\Delta\phi$, this matrix `grows' from the central row towards the first and last rows at the same time. 

A final speedup comes by noting that we are not limited to growing a chain from the center out to the edges, but we may instead set up multiple subchains. E.g., we can have 2 subchains which cover the ranges $\phi\in [0, \pi]$ and $\phi\in[\pi, 2\pi]$ that are then themselves grown from the center outwards. Indeed, one could take this a step further and completely parallelize the process to find \textit{all} the roots for \textit{all} the desired values of $\phi$ at the same time. Nothing is stopping us from finding the $n_r$ roots for $\phi=0$ at the same time as we find, e.g., the roots for $\phi=\pi/100, 2\pi/100, 3\pi/100,...$, since every set of roots for a particular value of $\phi$ is independent of the roots for another value of $\phi$. In practice however, this makes it more difficult to determine which roots form part of the same chains. We want to be able to clearly trace the critical curves as $\phi$ varies, not end up with a collection of randomly determined roots for various values of $\phi$; this necessitates stepping through values of $\Delta\phi$ to contextualize the locations of roots relative to each other. The problem ultimately becomes one of minimizing the runtime while also minimizing the number of subchains to calculate. We are content with our current implementation, though a deeper examination of the computational methods might provide an alternative ideal implementation.

\subsection{Runtime analyses} 

We have incorporated the methods discussed above into a CUDA program using double precision arithmetic. Figure~\ref{fig:time_vs_num_stars} shows the time required to calculate the whole critical curve network as a function of the number of microlenses $N_\star$, for various values of $m$ and for both circular and rectangular microlens fields. The runtime of the parallelized Aberth-Ehrlich method for a single value of $\phi$ has an asymptotic dependence that is $O(N_\star^2)$ \citep{2017GhiEA}. While this strictly holds true for calculating the initial roots, the time required goes down for subsequent values of $\phi$ due to the proximity of roots separated by $\Delta\phi$. This results in a roughly linear dependence on $N_\star$ when using the FMM \citep{1993A&A...268..501W} for low $N_\star$, though a quadratic runtime dependence on $N_\star$ eventually becomes apparent for large enough $N_\star\gtrsim10^4$ as seen in Figure~\ref{fig:time_vs_num_stars}.

\section{The number of caustic crossings}
\label{sec:calculating_ncc}

\subsection{Calculating the number of caustic crossings}

\begin{figure*}
    \centering
    \includegraphics[width=\textwidth]{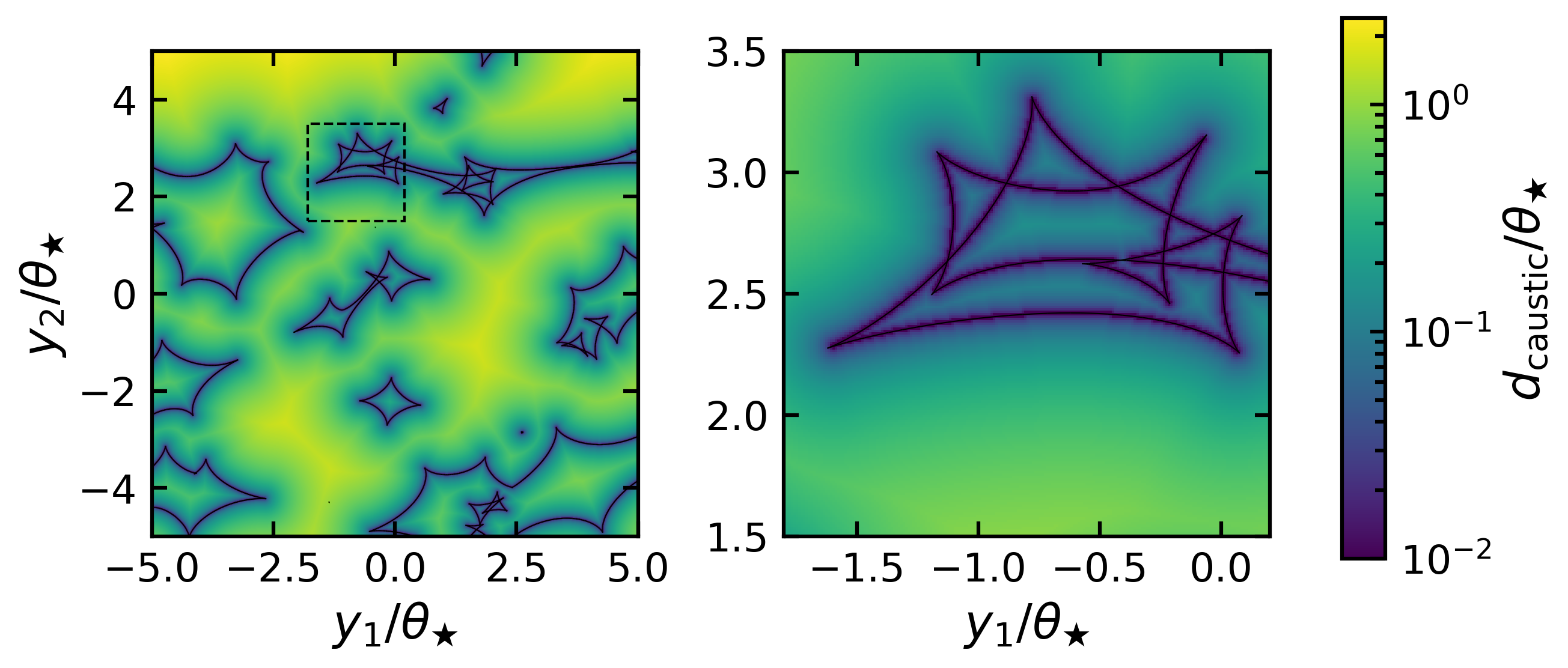}\\
    \includegraphics[width=\textwidth]{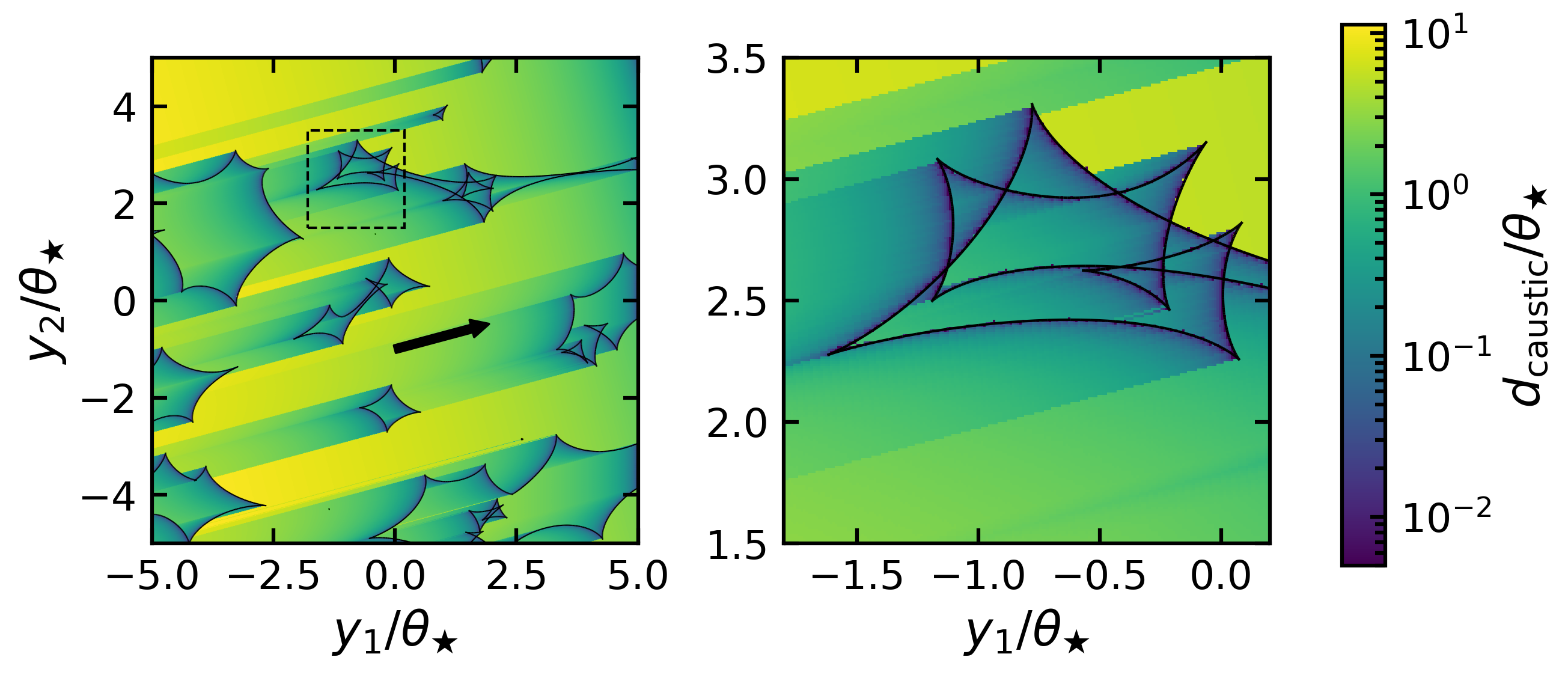}
    \caption{The distance $d_\text{caustic}$ from each (pixelated) source position to the nearest caustic for an expanding source (top) and for a moving source (bottom) traveling along the direction indicated by the black arrow. We note that edge effects must additionally be accounted for, as the border of the map is a boundary that can be hit in the distance algorithms despite not being a caustic.}
    \label{fig:d_caustic}
\end{figure*}

The chosen parametrization of $\phi$ traces the critical curves in a direction such that as $\phi$ increases, regions of positive parity are to the right of the critical curve and regions of negative parity are to the left; see Figure~\ref{fig:crit_curves}, as microlenses must always lie in regions of negative parity. Translating to the source plane, this means that the caustics are traced out with a clockwise orientation \citep{1990A&A...236..311W, 2015ApJ...806...63D}. 

While \citet{1992A&A...258..591W} and \citet{2003ApJ...583..575G} use the parametric representation of the caustics to calculate maps which give (essentially) the number of caustic crossings $N_\text{caustic crossings}$ as a function of position in the source plane, they do not comment on the numerical methods used and we must therefore implement our own. Since the caustics are oriented curves, winding numbers provide the number of caustic crossings. For every (pixelized) source location, we use a GPU version of \citeauthor{2001Sunday}'s \citeyearpar{2001Sunday} algorithm to calculate the winding number of the discretized caustic polygons around that point, efficiently creating a map that provides the number of caustic crossings\footnote{Slightly technical, but it turns out to be computationally faster to loop over the caustic segments and calculate which pixels are affected by a given segment, rather than looping over pixels and calculating which caustic segments affect a given pixel.}. We show in Figure~\ref{fig:ncc} the output map of $N_\text{caustic crossings}$, overlaid with the positions of the caustics. Although one could count the number of crossings by visual inspection (with some difficulty in dense regions), our ability to calculate them with a consistent algorithm is of some interest as detailed below.

\subsection{Applications}

By separating the source plane into regions with distinct integer values, various image processing algorithms can be used to determine distances between regions. We show in the top of Figure~\ref{fig:d_caustic} the distance from each pixel to the nearest caustic for an expanding source (e.g. a supernova), calculated from the map of $N_\text{caustic crossings}$ using a Euclidean distance transform. The bottom of Figure~\ref{fig:d_caustic} shows the distance to the nearest caustic for a source moving along a particular direction (e.g. a quasar). The former distance map can be used to constrain the ratio of supernova size to $\theta_\star$ given the presence or absence of caustic crossings, which can subsequently be used to infer the average stellar mass if supernova size is known \citep{2025arXiv250201728W}. The latter can be used to constrain the direction and velocity of quasars, or to act as a prior on the source position when fitting light curves that exhibit caustic crossings. 

\begin{figure}
    \centering
    \includegraphics[width=0.5\textwidth]{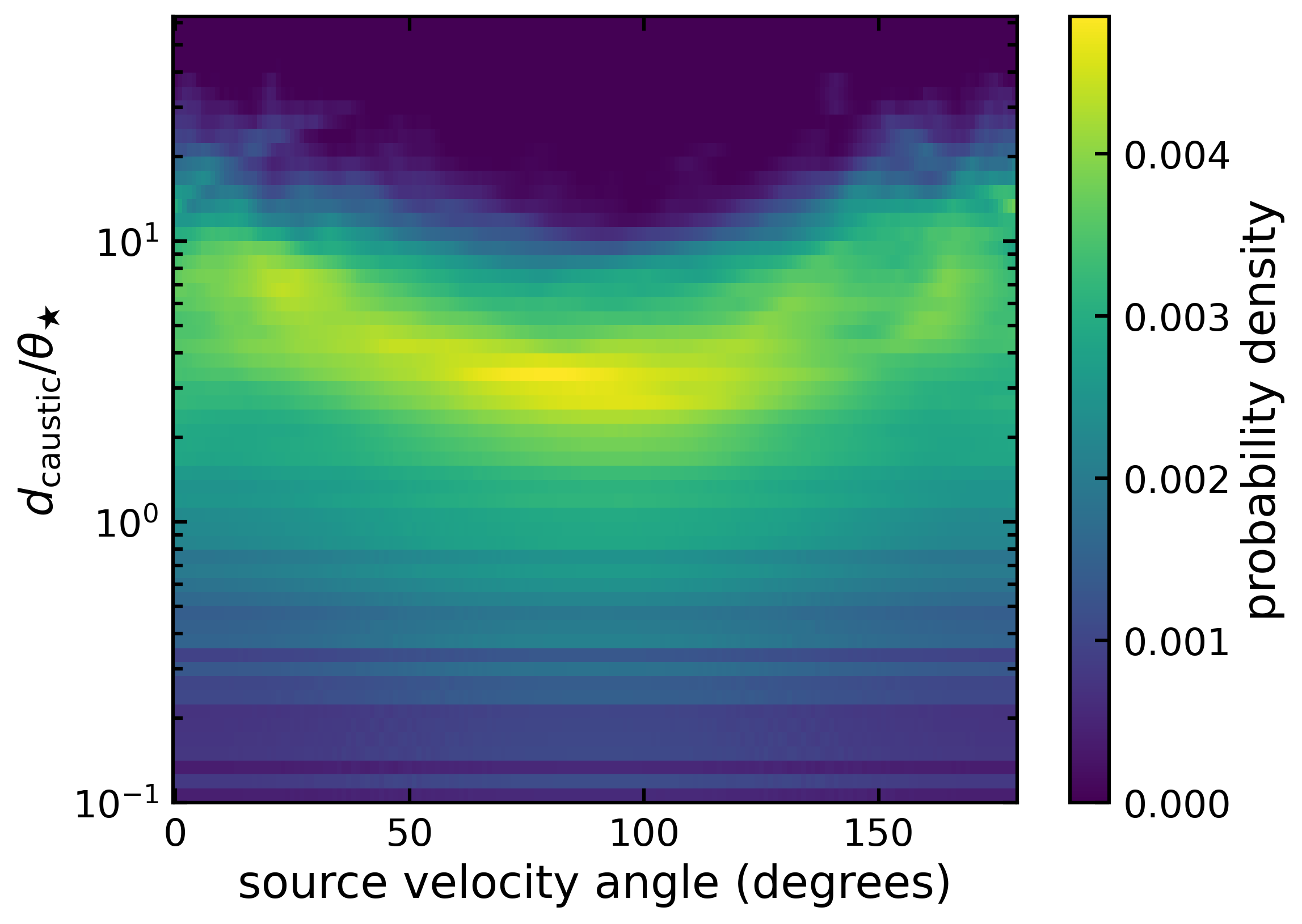}
    \caption{Sample probability distribution of the distance to the nearest caustic and the source velocity angle with respect to the positive $y_1$ axis. The distribution makes it clear that travel perpendicular to the caustics (90\textdegree) experiences caustic crossings more frequently than travel parallel to the caustics; an obvious well known fact, but one that this figure can numerically quantify without explicitly simulating light curves. We note that the horizontal stripes are artifacts of finite pixel size and rotation discretizations.}
    \label{fig:p_d_caustic_angle}
\end{figure}

We show in Figure~\ref{fig:p_d_caustic_angle} the probability distribution of the distance to the nearest caustic for a moving source as a function of the angle the source velocity makes with the $y_1$ axis. The distribution highlights the well-known fact that movement parallel to the caustics (in the $y_1$ direction, when the angle is 0\textdegree) typically experiences longer periods between caustic crossings than movement perpendicular to the caustics (in the $y_2$ direction, when the angle is 90\textdegree) -- in this particular case by a factor of $\sim3$.

\begin{figure}
    \centering
    \includegraphics[width=0.5\textwidth]{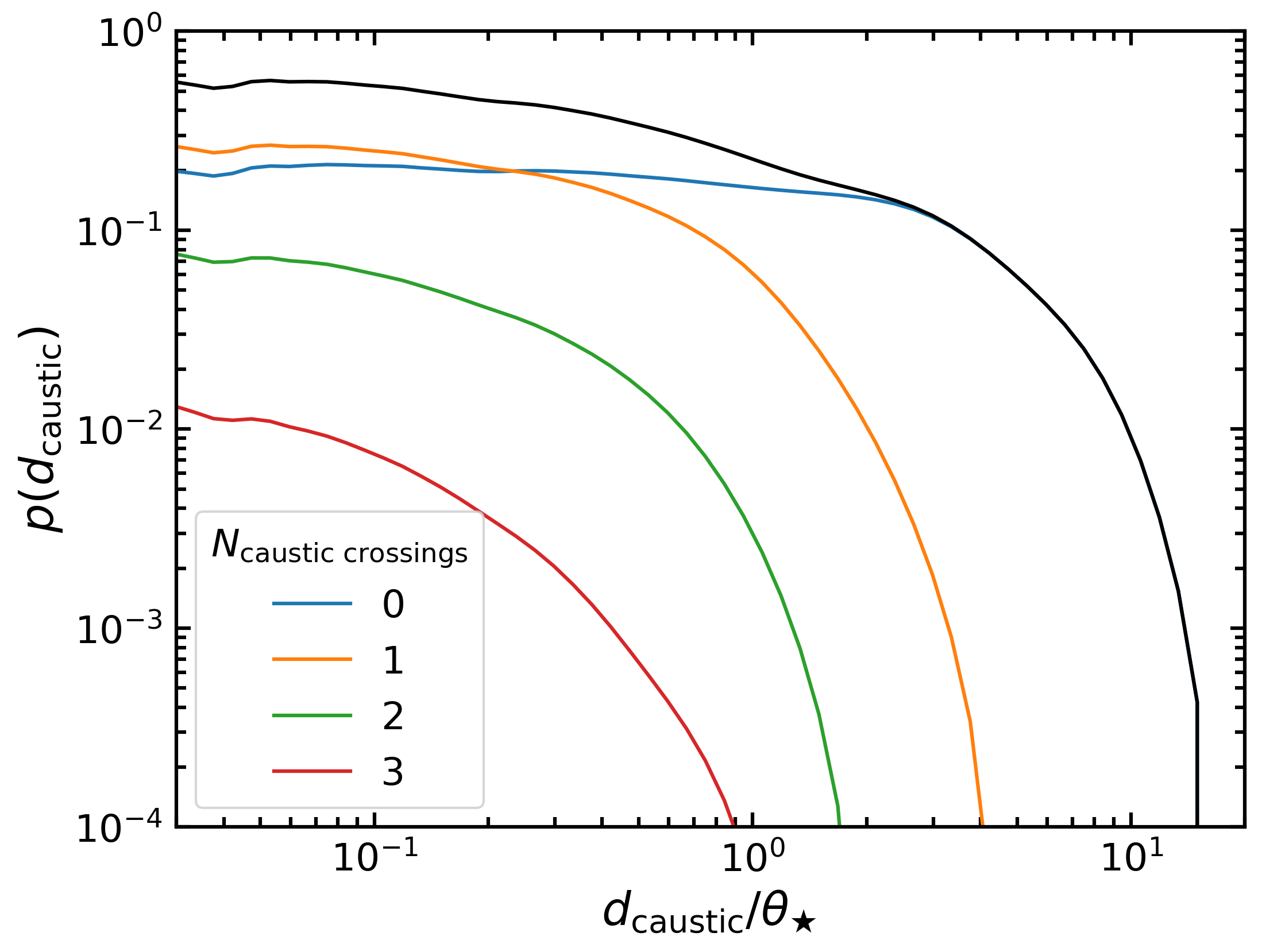}
    \caption{Sample probability distribution of the distance to the nearest caustic for a source traveling along the direction $y_2$ (i.e. a vertical cut of Figure~\ref{fig:p_d_caustic_angle} for a source velocity angle of 90\textdegree). The total distribution (solid black line) can be decomposed into subdistributions based on where the source is located in the map of $N_\text{caustic crossings}$ (Figure~\ref{fig:ncc}). The distance between caustic crossing events can be used to pin down where the source is located on the magnification map; e.g., a source that has traveled $5\theta_\star$ \textit{must} be located outside the microcaustics ($N_\text{caustic crossings}=0$).}
    \label{fig:p_d_caustic_quasar}
\end{figure}

We can additionally determine the distance between caustics as a function of where the source happens to lie in the map of $N_\text{caustic crossings}$. Figure~\ref{fig:p_d_caustic_quasar} shows an example for movement along the $y_2$ axis. The sharp cutoffs in the distributions make it obvious that a moving source can only lie within the caustics for a limited period of time. A measurement of the time between subsequent caustic crossing events can therefore be used to constrain regions of the source plane that the source might have moved through, though one most account for the degeneracy between $\theta_\star$ and source velocity. 

\section{Conclusions}
\label{sec:conclusions}

In anticipation of a drastic increase in the number of gravitationally lensed supernovae and quasars from upcoming surveys, it is vital to improve upon techniques for analyzing observations of microlensed systems. Incorporating additional information regarding the locations of microlensing caustics into analyses of light curves or HMEs on top of typical approaches that utilize magnification maps can improve constraints on astrophysical and cosmological parameters of interest.

We have presented in this work a GPU code to calculate the critical curves and caustics of the microlenses. We use the Aberth-Ehrlich method \citep{1967Ehrlich, 1973Aberth}, a cubically convergent algorithm that allows for parallelization of finding the roots of polynomials. Following \citet{2017GhiEA}, we implement the algorithm on a GPU to efficiently solve the very high degree polynomials required for finding the locations of the critical curves of large numbers of microlenses. The Aberth-Ehrlich method has also recently been used in the context of exoplanet microlensing to efficiently solve the 5th order polynomial that provides the microimage positions of binary lenses on both CPUs \citep{2022MNRAS.514.4379F} and GPUs \citep{2025ApJS..276...40W, 2025AJ....169..170R}. While our focus is on the vast caustic network typical for quasar and supernovae microlensing, there is no reason our code could not also be used to locate the critical curves and caustics of exoplanet microlensing systems with an arbitrary number of planets.

Since the caustics are oriented curves, they can be used to calculate a pixelized map of the number of caustic crossings as a function of source position. We have shown how this map can be used to calculate the distances between caustics for expanding or moving sources. The resulting maps and probability distributions of distances can be used to constrain source location and velocity within magnification maps, helping to improve the efficiency of microlensing calculations and constrain microlensing (de)magnifications as well \citep{2024MNRAS.531.4349W}. Using caustic crossing information to also constrain the microlens mass simultaneously with the stellar mass fraction has been examined in \citet{2025arXiv250201728W}. Both analyses in the above works required the computational methods presented here in order to generate likelihoods. Improving and incorporating those methods into analyses for future lensed supernovae will allow us to make astrophysical constraints on various parameters related to stellar mass in the lens galaxy, in addition to cosmological constraints from time delay cosmography utilizing the standardizable candle nature of Type Ia supernovae. 

Another application of determining the locations of the critical curves discussed in \citet{1990A&A...236..311W} is the ability to calculate distributions of the parameter that governs (along with the source size) the maximum achievable magnification during a caustic crossing event. This parameter is a combination of derivatives of the lensing potential evaluated along the critical curve. While \citet{1990A&A...236..311W} and \citet{1998JKAS...31...27L} provide distributions for some specific systems and for a range of convergences with no external shear, a full examination of the parameter and its dependence on the macromodel convergence and shear is now more numerically feasible. Calculations of the likelihood of this parameter will be important in analyzing HMEs when larger numbers of them begin to be observed. In addition, some authors \citep{2011MNRAS.417..541A} have proposed using higher order derivatives of the lensing potential to more fully characterize caustic crossing events; distributions of such additional parameters have previously been unexplored. 

Of additional interest is the following. \citet{2003ApJ...583..575G} and \citet{2011MNRAS.411.1671S} both discuss how the microlensing magnification probability distribution can be decomposed based on regions with varying number of microminima; the importance of the microminima is a phenomenon also noted in \citet{1992A&A...258..591W}. In brief, the magnification maps mentioned in Section~\ref{sec:theory} are combined with maps of the number of caustic crossings from Section~\ref{sec:calculating_ncc} to decompose the magnification distribution. The subdistributions all appear self-similar (see, e.g., Figure 4 from \citealp{2003ApJ...583..575G}, Figure 3 from \citealp{2011MNRAS.411.1671S}, or Figure 1 of \citealp{2024MNRAS.531.4349W}), suggesting that their sum can be understood from the individual parts. In addition, the distributions appear to change smoothly through microlensing parameter space, suggesting that there may be a functional form which can be used to derive magnification probabilities analytically rather than numerically -- a possibility which would be of great interest for modeling microlensed systems. To date, only a limited number of simulations have been performed to study and characterize the dependence of the magnification probability distribution on the number of microminima, likely due to the time intensive nature of rayshooting and finding the caustics when those studies were performed. With the more recent advent and rise of GPU computing and the improvements presented here, the time is ripe for a more detailed study which we hope to perform in the future. 

\section*{Acknowledgements}

The author would like to thank Robert Schmidt for (perhaps unknowingly) introducing him to C implementations of Hans Witt's caustic finding code. He would like to thank Aksel Alpay for an extremely enlightening presentation about GPU computing (applied to ray shooting) that prompted some investigations which ultimately led to this paper. He would also like to thank Joachim Wambsganss, and the rest of the ARI microlensing group, for their hospitality during his 2018-2019 stay.  

Numerical computations were done on the Sciama High Performance Compute (HPC) cluster which is supported by the ICG, SEPNet, and the University of Portsmouth. 

Parts of this work were supported by the Deutsch-Amerikanische Fulbright-Kommission. This work has also received funding from the European Research Council (ERC) under the European Union’s Horizon 2020 research and innovation programme (LensEra: grant agreement No. 945536). For the purpose of open access, the authors have applied a Creative Commons Attribution (CC BY) license to any Author Accepted Manuscript version arising.

\section*{Data Availability}

Data from this work can be made available upon reasonable request to the corresponding author. The code developed is publicly available and linked in the abstract. Bugs can be raised as issues on github or reported to the author via email. Running the code requires an NVIDIA graphics card. The NVIDIA CUDA compiler \texttt{nvcc} is required to compile the code, as well as a C++20 compliant compiler. Precompiled libraries created using the GNU compiler v11.2.0 and the CUDA compiler v12.4 are also provided, which should work on Linux distributions that have GLIBC >= 2.31 and GLIBCXX >= 3.4.29.



\bibliographystyle{mnras}
\bibliography{bibliography.bib} 

\begin{thebibliography}{}
\makeatletter
\relax
\def\mn@urlcharsother{\let\do\@makeother \do\$\do\&\do\#\do\^\do\_\do\%\do\~}
\def\mn@doi{\begingroup\mn@urlcharsother \@ifnextchar [ {\mn@doi@} {\mn@doi@[]}}
\def\mn@doi@[#1]#2{\def\@tempa{#1}\ifx\@tempa\@empty \href {http://dx.doi.org/#2} {doi:#2}\else \href {http://dx.doi.org/#2} {#1}\fi \endgroup}
\def\mn@eprint#1#2{\mn@eprint@#1:#2::\@nil}
\def\mn@eprint@arXiv#1{\href {http://arxiv.org/abs/#1} {{\tt arXiv:#1}}}
\def\mn@eprint@dblp#1{\href {http://dblp.uni-trier.de/rec/bibtex/#1.xml} {dblp:#1}}
\def\mn@eprint@#1:#2:#3:#4\@nil{\def\@tempa {#1}\def\@tempb {#2}\def\@tempc {#3}\ifx \@tempc \@empty \let \@tempc \@tempb \let \@tempb \@tempa \fi \ifx \@tempb \@empty \def\@tempb {arXiv}\fi \@ifundefined {mn@eprint@\@tempb}{\@tempb:\@tempc}{\expandafter \expandafter \csname mn@eprint@\@tempb\endcsname \expandafter{\@tempc}}}

\bibitem[\protect\citeauthoryear{{Aberth}}{{Aberth}}{1973}]{1973Aberth}
{Aberth} O.,  1973, Mathematics of Computation, 27, 339–344

\bibitem[\protect\citeauthoryear{{Alexandrov} \& {Zhdanov}}{{Alexandrov} \& {Zhdanov}}{2011}]{2011MNRAS.417..541A}
{Alexandrov} A.~N.,  {Zhdanov} V.~I.,  2011, \mn@doi [\mnras] {10.1111/j.1365-2966.2011.19296.x}, \href {https://ui.adsabs.harvard.edu/#abs/2011MNRAS.417..541A} {417, 541}

\bibitem[\protect\citeauthoryear{{Alpay}}{{Alpay}}{2019}]{teralens}
{Alpay} A.,  2019, \url {https://github.com/illuhad/teralens}

\bibitem[\protect\citeauthoryear{{Bate}, {Fluke}, {Barsdell}, {Garsden}  \& {Lewis}}{{Bate} et~al.}{2010}]{2010NewA...15..726B}
{Bate} N.~F.,  {Fluke} C.~J.,  {Barsdell} B.~R.,  {Garsden} H.,   {Lewis} G.~F.,  2010, \mn@doi [\na] {10.1016/j.newast.2010.05.008}, \href {https://ui.adsabs.harvard.edu/abs/2010NewA...15..726B} {15, 726}

\bibitem[\protect\citeauthoryear{{Best}, {Fagin}, {Vernardos}  \& {O'Dowd}}{{Best} et~al.}{2024}]{2024MNRAS.531.1095B}
{Best} H.,  {Fagin} J.,  {Vernardos} G.,   {O'Dowd} M.,  2024, \mn@doi [\mnras] {10.1093/mnras/stae1182}, \href {https://ui.adsabs.harvard.edu/abs/2024MNRAS.531.1095B} {531, 1095}

\bibitem[\protect\citeauthoryear{{Bourassa} \& {Kantowski}}{{Bourassa} \& {Kantowski}}{1975}]{1975ApJ...195...13B}
{Bourassa} R.~R.,  {Kantowski} R.,  1975, \mn@doi [\apj] {10.1086/153300}, \href {https://ui.adsabs.harvard.edu/abs/1975ApJ...195...13B} {195, 13}

\bibitem[\protect\citeauthoryear{{Bourassa}, {Kantowski}  \& {Norton}}{{Bourassa} et~al.}{1973}]{1973ApJ...185..747B}
{Bourassa} R.~R.,  {Kantowski} R.,   {Norton} T.~D.,  1973, \mn@doi [\apj] {10.1086/152452}, \href {https://ui.adsabs.harvard.edu/abs/1973ApJ...185..747B} {185, 747}

\bibitem[\protect\citeauthoryear{{Chartas}, {Kochanek}, {Dai}, {Poindexter}  \& {Garmire}}{{Chartas} et~al.}{2009}]{2009ApJ...693..174C}
{Chartas} G.,  {Kochanek} C.~S.,  {Dai} X.,  {Poindexter} S.,   {Garmire} G.,  2009, \mn@doi [\apj] {10.1088/0004-637X/693/1/174}, \href {https://ui.adsabs.harvard.edu/abs/2009ApJ...693..174C} {693, 174}

\bibitem[\protect\citeauthoryear{{Dan{\v{e}}k} \& {Heyrovsk{\'y}}}{{Dan{\v{e}}k} \& {Heyrovsk{\'y}}}{2015}]{2015ApJ...806...63D}
{Dan{\v{e}}k} K.,  {Heyrovsk{\'y}} D.,  2015, \mn@doi [\apj] {10.1088/0004-637X/806/1/63}, \href {https://ui.adsabs.harvard.edu/abs/2015ApJ...806...63D} {806, 63}

\bibitem[\protect\citeauthoryear{{Ehrlich}}{{Ehrlich}}{1967}]{1967Ehrlich}
{Ehrlich} L.~W.,  1967, Communications of the ACM, 10, 107

\bibitem[\protect\citeauthoryear{{Fatheddin} \& {Sajadian}}{{Fatheddin} \& {Sajadian}}{2022}]{2022MNRAS.514.4379F}
{Fatheddin} H.,  {Sajadian} S.,  2022, \mn@doi [\mnras] {10.1093/mnras/stac1565}, \href {https://ui.adsabs.harvard.edu/abs/2022MNRAS.514.4379F} {514, 4379}

\bibitem[\protect\citeauthoryear{{Ghidouche}, {Sider}, {Couturier}  \& {Guyeux}}{{Ghidouche} et~al.}{2017}]{2017GhiEA}
{Ghidouche} K.,  {Sider} A.,  {Couturier} R.,   {Guyeux} C.,  2017, Journal of Computational Science, 18, 46

\bibitem[\protect\citeauthoryear{{Granot}, {Schechter}  \& {Wambsganss}}{{Granot} et~al.}{2003}]{2003ApJ...583..575G}
{Granot} J.,  {Schechter} P.~L.,   {Wambsganss} J.,  2003, \mn@doi [\apj] {10.1086/345447}, \href {https://ui.adsabs.harvard.edu/abs/2003ApJ...583..575G} {583, 575}

\bibitem[\protect\citeauthoryear{{Greengard} \& {Rokhlin}}{{Greengard} \& {Rokhlin}}{1987}]{1987JCoPh..73..325G}
{Greengard} L.,  {Rokhlin} V.,  1987, \mn@doi [Journal of Computational Physics] {10.1016/0021-9991(87)90140-9}, \href {https://ui.adsabs.harvard.edu/abs/1987JCoPh..73..325G} {73, 325}

\bibitem[\protect\citeauthoryear{{Hundertmark}}{{Hundertmark}}{2011}]{2011PhDT.......335H}
{Hundertmark} M. P.~G.,  2011, PhD thesis, Georg August University of Gottingen, Germany

\bibitem[\protect\citeauthoryear{{Jim{\'e}nez-Vicente} \& {Mediavilla}}{{Jim{\'e}nez-Vicente} \& {Mediavilla}}{2022}]{2022ApJ...941...80J}
{Jim{\'e}nez-Vicente} J.,  {Mediavilla} E.,  2022, \mn@doi [\apj] {10.3847/1538-4357/ac9e59}, \href {https://ui.adsabs.harvard.edu/abs/2022ApJ...941...80J} {941, 80}

\bibitem[\protect\citeauthoryear{{Kayser}, {Refsdal}  \& {Stabell}}{{Kayser} et~al.}{1986}]{1986A&A...166...36K}
{Kayser} R.,  {Refsdal} S.,   {Stabell} R.,  1986, \aap, \href {https://ui.adsabs.harvard.edu/abs/1986A&A...166...36K} {166, 36}

\bibitem[\protect\citeauthoryear{{Kochanek}}{{Kochanek}}{2004}]{2004ApJ...605...58K}
{Kochanek} C.~S.,  2004, \mn@doi [\apj] {10.1086/382180}, \href {https://ui.adsabs.harvard.edu/abs/2004ApJ...605...58K} {605, 58}

\bibitem[\protect\citeauthoryear{{Lee}, {Chang}  \& {Kim}}{{Lee} et~al.}{1998}]{1998JKAS...31...27L}
{Lee} D.~W.,  {Chang} K.~A.,   {Kim} S.~J.,  1998, Journal of Korean Astronomical Society, \href {https://ui.adsabs.harvard.edu/#abs/1998JKAS...31...27L} {31, 27}

\bibitem[\protect\citeauthoryear{{Lewis}, {Miralda-Escude}, {Richardson}  \& {Wambsganss}}{{Lewis} et~al.}{1993}]{1993MNRAS.261..647L}
{Lewis} G.~F.,  {Miralda-Escude} J.,  {Richardson} D.~C.,   {Wambsganss} J.,  1993, \mn@doi [\mnras] {10.1093/mnras/261.3.647}, \href {https://ui.adsabs.harvard.edu/abs/1993MNRAS.261..647L} {261, 647}

\bibitem[\protect\citeauthoryear{{Mediavilla}, {Mu{\~n}oz}, {Lopez}, {Mediavilla}, {Abajas}, {Gonzalez-Morcillo}  \& {Gil-Merino}}{{Mediavilla} et~al.}{2006}]{2006ApJ...653..942M}
{Mediavilla} E.,  {Mu{\~n}oz} J.~A.,  {Lopez} P.,  {Mediavilla} T.,  {Abajas} C.,  {Gonzalez-Morcillo} C.,   {Gil-Merino} R.,  2006, \mn@doi [\apj] {10.1086/508796}, \href {https://ui.adsabs.harvard.edu/abs/2006ApJ...653..942M} {653, 942}

\bibitem[\protect\citeauthoryear{{Mediavilla}, {Mediavilla}, {Mu{\~n}oz}, {Ariza}, {Lopez}, {Gonzalez-Morcillo}  \& {Jimenez-Vicente}}{{Mediavilla} et~al.}{2011}]{2011ApJ...741...42M}
{Mediavilla} E.,  {Mediavilla} T.,  {Mu{\~n}oz} J.~A.,  {Ariza} O.,  {Lopez} P.,  {Gonzalez-Morcillo} C.,   {Jimenez-Vicente} J.,  2011, \mn@doi [\apj] {10.1088/0004-637X/741/1/42}, \href {https://ui.adsabs.harvard.edu/abs/2011ApJ...741...42M} {741, 42}

\bibitem[\protect\citeauthoryear{{Natarajan}, {Williams}, {Brada{\v{c}}}, {Grillo}, {Ghosh}, {Sharon}  \& {Wagner}}{{Natarajan} et~al.}{2024}]{2024SSRv..220...19N}
{Natarajan} P.,  {Williams} L.~L.~R.,  {Brada{\v{c}}} M.,  {Grillo} C.,  {Ghosh} A.,  {Sharon} K.,   {Wagner} J.,  2024, \mn@doi [\ssr] {10.1007/s11214-024-01051-8}, \href {https://ui.adsabs.harvard.edu/abs/2024SSRv..220...19N} {220, 19}

\bibitem[\protect\citeauthoryear{{Neira}, {Anguita}  \& {Vernardos}}{{Neira} et~al.}{2020}]{2020MNRAS.495..544N}
{Neira} F.,  {Anguita} T.,   {Vernardos} G.,  2020, \mn@doi [\mnras] {10.1093/mnras/staa1208}, \href {https://ui.adsabs.harvard.edu/abs/2020MNRAS.495..544N} {495, 544}

\bibitem[\protect\citeauthoryear{{Paczynski}}{{Paczynski}}{1986}]{1986ApJ...301..503P}
{Paczynski} B.,  1986, \mn@doi [\apj] {10.1086/163919}, \href {https://ui.adsabs.harvard.edu/abs/1986ApJ...301..503P} {301, 503}

\bibitem[\protect\citeauthoryear{{Paic}, {Vernardos}, {Sluse}, {Millon}, {Courbin}, {Chan}  \& {Bonvin}}{{Paic} et~al.}{2022}]{2022A&A...659A..21P}
{Paic} E.,  {Vernardos} G.,  {Sluse} D.,  {Millon} M.,  {Courbin} F.,  {Chan} J.~H.,   {Bonvin} V.,  2022, \mn@doi [\aap] {10.1051/0004-6361/202141808}, \href {https://ui.adsabs.harvard.edu/abs/2022A&A...659A..21P} {659, A21}

\bibitem[\protect\citeauthoryear{{Ren} \& {Zhu}}{{Ren} \& {Zhu}}{2025}]{2025AJ....169..170R}
{Ren} H.,  {Zhu} W.,  2025, \mn@doi [\aj] {10.3847/1538-3881/adb1b2}, \href {https://ui.adsabs.harvard.edu/abs/2025AJ....169..170R} {169, 170}

\bibitem[\protect\citeauthoryear{{Saha} \& {Williams}}{{Saha} \& {Williams}}{2011}]{2011MNRAS.411.1671S}
{Saha} P.,  {Williams} L. L.~R.,  2011, \mn@doi [\mnras] {10.1111/j.1365-2966.2010.17797.x}, \href {https://ui.adsabs.harvard.edu/abs/2011MNRAS.411.1671S} {411, 1671}

\bibitem[\protect\citeauthoryear{{Saha}, {Sluse}, {Wagner}  \& {Williams}}{{Saha} et~al.}{2024}]{2024SSRv..220...12S}
{Saha} P.,  {Sluse} D.,  {Wagner} J.,   {Williams} L. L.~R.,  2024, \mn@doi [\ssr] {10.1007/s11214-024-01041-w}, \href {https://ui.adsabs.harvard.edu/abs/2024SSRv..220...12S} {220, 12}

\bibitem[\protect\citeauthoryear{{Schneider} \& {Weiss}}{{Schneider} \& {Weiss}}{1986}]{1986A&A...164..237S}
{Schneider} P.,  {Weiss} A.,  1986, \aap, \href {https://ui.adsabs.harvard.edu/abs/1986A&A...164..237S} {164, 237}

\bibitem[\protect\citeauthoryear{{Shajib} et~al.,}{{Shajib} et~al.}{2022}]{2022arXiv221010790S}
{Shajib} A.~J.,  et~al., 2022, \mn@doi [arXiv e-prints] {10.48550/arXiv.2210.10790}, \href {https://ui.adsabs.harvard.edu/abs/2022arXiv221010790S} {p. arXiv:2210.10790}

\bibitem[\protect\citeauthoryear{{Shalyapin}, {Gil-Merino}  \& {Goicoechea}}{{Shalyapin} et~al.}{2021}]{2021A&A...653A.121S}
{Shalyapin} V.~N.,  {Gil-Merino} R.,   {Goicoechea} L.~J.,  2021, \mn@doi [\aap] {10.1051/0004-6361/202140527}, \href {https://ui.adsabs.harvard.edu/abs/2021A&A...653A.121S} {653, A121}

\bibitem[\protect\citeauthoryear{{Sunday}}{{Sunday}}{2001}]{2001Sunday}
{Sunday} D.,  2001, Inclusion of a Point in a Polygon, \url {https://web.archive.org/web/20130126163405/http://geomalgorithms.com/a03-_inclusion.html}

\bibitem[\protect\citeauthoryear{{Suyu}, {Goobar}, {Collett}, {More}  \& {Vernardos}}{{Suyu} et~al.}{2024}]{2024SSRv..220...13S}
{Suyu} S.~H.,  {Goobar} A.,  {Collett} T.,  {More} A.,   {Vernardos} G.,  2024, \mn@doi [\ssr] {10.1007/s11214-024-01044-7}, \href {https://ui.adsabs.harvard.edu/abs/2024SSRv..220...13S} {220, 13}

\bibitem[\protect\citeauthoryear{{Thompson}, {Fluke}, {Barnes}  \& {Barsdell}}{{Thompson} et~al.}{2010}]{2010NewA...15...16T}
{Thompson} A.~C.,  {Fluke} C.~J.,  {Barnes} D.~G.,   {Barsdell} B.~R.,  2010, \mn@doi [\na] {10.1016/j.newast.2009.05.010}, \href {https://ui.adsabs.harvard.edu/abs/2010NewA...15...16T} {15, 16}

\bibitem[\protect\citeauthoryear{{Vernardos} \& {Tsagkatakis}}{{Vernardos} \& {Tsagkatakis}}{2019}]{2019MNRAS.486.1944V}
{Vernardos} G.,  {Tsagkatakis} G.,  2019, \mn@doi [\mnras] {10.1093/mnras/stz868}, \href {https://ui.adsabs.harvard.edu/abs/2019MNRAS.486.1944V} {486, 1944}

\bibitem[\protect\citeauthoryear{{Vernardos} et~al.,}{{Vernardos} et~al.}{2024}]{2024SSRv..220...14V}
{Vernardos} G.,  et~al., 2024, \mn@doi [\ssr] {10.1007/s11214-024-01043-8}, \href {https://ui.adsabs.harvard.edu/abs/2024SSRv..220...14V} {220, 14}

\bibitem[\protect\citeauthoryear{{Walsh}, {Carswell}  \& {Weymann}}{{Walsh} et~al.}{1979}]{1979Natur.279..381W}
{Walsh} D.,  {Carswell} R.~F.,   {Weymann} R.~J.,  1979, \mn@doi [Nature] {10.1038/279381a0}, \href {https://ui.adsabs.harvard.edu/abs/1979Natur.279..381W} {279, 381}

\bibitem[\protect\citeauthoryear{{Wambsganss}}{{Wambsganss}}{1990}]{1990PhDT.......180W}
{Wambsganss} J.,  1990, PhD thesis, -

\bibitem[\protect\citeauthoryear{{Wambsganss}, {Witt}  \& {Schneider}}{{Wambsganss} et~al.}{1992}]{1992A&A...258..591W}
{Wambsganss} J.,  {Witt} H.~J.,   {Schneider} P.,  1992, \aap, \href {https://ui.adsabs.harvard.edu/abs/1992A&A...258..591W} {258, 591}

\bibitem[\protect\citeauthoryear{{Wang}, {Wang}  \& {Dong}}{{Wang} et~al.}{2025}]{2025ApJS..276...40W}
{Wang} S.,  {Wang} L.,   {Dong} S.,  2025, \mn@doi [\apjs] {10.3847/1538-4365/ad9b8d}, \href {https://ui.adsabs.harvard.edu/abs/2025ApJS..276...40W} {276, 40}

\bibitem[\protect\citeauthoryear{{Weisenbach}, {Collett}, {de Murieta}, {Krawczyk}, {Vernardos}, {Enzi}  \& {Lundgren}}{{Weisenbach} et~al.}{2024}]{2024MNRAS.531.4349W}
{Weisenbach} L.,  {Collett} T.,  {de Murieta} A.~S.,  {Krawczyk} C.,  {Vernardos} G.,  {Enzi} W.,   {Lundgren} A.,  2024, \mn@doi [\mnras] {10.1093/mnras/stae1396}, \href {https://ui.adsabs.harvard.edu/abs/2024MNRAS.531.4349W} {531, 4349}

\bibitem[\protect\citeauthoryear{{Weisenbach}, {Collett}, {Enzi}, {Oldham}  \& {Sainz de Murieta}}{{Weisenbach} et~al.}{2025}]{2025arXiv250201728W}
{Weisenbach} L.,  {Collett} T.,  {Enzi} W.,  {Oldham} L.,   {Sainz de Murieta} A.,  2025, \mn@doi [arXiv e-prints] {10.48550/arXiv.2502.01728}, \href {https://ui.adsabs.harvard.edu/abs/2025arXiv250201728W} {p. arXiv:2502.01728}

\bibitem[\protect\citeauthoryear{{Witt}}{{Witt}}{1990}]{1990A&A...236..311W}
{Witt} H.~J.,  1990, \aap, \href {https://ui.adsabs.harvard.edu/abs/1990A&A...236..311W} {236, 311}

\bibitem[\protect\citeauthoryear{{Witt}}{{Witt}}{1993}]{1993ApJ...403..530W}
{Witt} H.~J.,  1993, \mn@doi [\apj] {10.1086/172223}, \href {https://ui.adsabs.harvard.edu/abs/1993ApJ...403..530W} {403, 530}

\bibitem[\protect\citeauthoryear{{Witt}, {Kayser}  \& {Refsdal}}{{Witt} et~al.}{1993}]{1993A&A...268..501W}
{Witt} H.~J.,  {Kayser} R.,   {Refsdal} S.,  1993, \aap, \href {https://ui.adsabs.harvard.edu/abs/1993A&A...268..501W} {268, 501}

\bibitem[\protect\citeauthoryear{{Wyithe} \& {Loeb}}{{Wyithe} \& {Loeb}}{2002}]{2002ApJ...577..615W}
{Wyithe} J. S.~B.,  {Loeb} A.,  2002, \mn@doi [\apj] {10.1086/342121}, \href {https://ui.adsabs.harvard.edu/abs/2002ApJ...577..615W} {577, 615}

\bibitem[\protect\citeauthoryear{{Wyithe} \& {Turner}}{{Wyithe} \& {Turner}}{2001}]{2001MNRAS.320...21W}
{Wyithe} J.~S.~B.,  {Turner} E.~L.,  2001, \mn@doi [\mnras] {10.1046/j.1365-8711.2001.03917.x}, \href {https://ui.adsabs.harvard.edu/abs/2001MNRAS.320...21W} {320, 21}

\bibitem[\protect\citeauthoryear{{Young}}{{Young}}{1981}]{1981ApJ...244..756Y}
{Young} P.,  1981, \mn@doi [\apj] {10.1086/158752}, \href {https://ui.adsabs.harvard.edu/abs/1981ApJ...244..756Y} {244, 756}

\bibitem[\protect\citeauthoryear{{Zheng}, {Chen}, {Li}  \& {Chen}}{{Zheng} et~al.}{2022}]{2022ApJ...931..114Z}
{Zheng} W.,  {Chen} X.,  {Li} G.,   {Chen} H.-Z.,  2022, \mn@doi [\apj] {10.3847/1538-4357/ac68ea}, \href {https://ui.adsabs.harvard.edu/abs/2022ApJ...931..114Z} {931, 114}

\makeatother
\end{thebibliography}




\appendix

\section{Removing discontinuities associated with \texorpdfstring{$\alpha_s$}{α\textunderscore s}}
\label{app:removing_discontinuities}

Let us consider a field of microlenses distributed in a circle of radius $R_\star$. In this case, \begin{equation}
    \alpha_s(z) = \begin{cases}
                -\kappa_\star z,& |z| \leq R_\star\\
                \frac{-\kappa_\star R_\star^2}{\overline{z}},& |z| > R_\star
            \end{cases}
\end{equation} and Equation~\eqref{eq:parametric_cc} becomes

\begin{equation}
    \begin{aligned}
    \gamma&+\theta_\star^{2}\sum_{i=1}^{N_\star}\frac{m_{i}}{(z-z_i)^2} - \frac{\kappa_\star R_\star^2}{z^2}\left(1 - H(R_\star - |z|)\right)\\
    &- \left(1 - \kappa + \kappa_\star H(R_\star - |z|)\right)e^{-i\phi}=0.
    \end{aligned}
    \label{eq:parametric_cc_circle}
\end{equation} where $H(x)$ is the Heaviside step function. 

Let us consider the two different regions for $z$: either inside ($H(R_\star - |z|)=1$) or outside ($H(R_\star-|z|)=0$) the circle of microlenses. In each case, for a particular value of $\phi$, Equation~\eqref{eq:parametric_cc_circle} can be transformed into a polynomial \citep{1990A&A...236..311W}.

Discontinuities in Equation~\eqref{eq:parametric_cc_circle} however means that we cannot smoothly trace these curves across the boundary of the microlensing field. We can get around the problem by expanding our disk of smooth matter, making it instead an infinite sheet such that $\alpha_s(z)=-\kappa_\star z$ everywhere. Inside the radius $R_\star$, nothing has changed -- the critical curves will be precisely where they are supposed to be. Outside of $R_\star$, the critical curves that we calculate will be incorrect; however, we are not in general interested in these areas\footnote{Strictly speaking, this is not true; we should still be concerned about critical curves outside the region of microlenses. Under this approximation of an infinite compensating mass sheet, systems which have $1-(\kappa-\kappa_\star)\pm\gamma\approx 0$ have caustics passing near the origin which come from critical curves very far from the origin, as the behavior far from the origin is equivalent to a Chang-Refsdal lens for a single microlens that has the total microlens mass and is embedded in a macromodel with near infinite magnification. This introduces complications when calculating the number of caustic crossings due to spurious caustics in the region of interest that come from critical curves outside the applicable microlens region.}. 

For a rectangular field of microlenses, $\alpha_s(z)$ can be found in \inprept{Weisenbach}, from which we can find
\begin{equation}
    \frac{\partial \alpha_s(z)}{\partial z} = -\kappa_\star B_{(c_1,c_2)}(z)
\end{equation} and

\begin{align}
    \begin{split}
    \frac{\partial \alpha_s(z)}{\partial \overline{z}} 
    =&\frac{-\kappa_\star i}{\pi}\Big[ -\log(c-\overline{z})+\log(\overline{c}-\overline{z})\\
    & -\log(-c-\overline{z})+\log(-\overline{c}-\overline{z})\Big]\\
    &-\kappa_\star B_{(c_1,c_2)}(z)
    \end{split}
\end{align} where $B_{(a,b)}$ is a two dimensional boxcar function \begin{align*}
    B_{(a,b)}(z) = \begin{cases}
                1,& -a \leq \operatorname{Re}(z) \leq a,\ -b \leq \operatorname{Im}(z) \leq b\\
                0,& \text{everywhere else}
            \end{cases}
\end{align*} and $c=c_1 + ic_2$ is the corner of the rectangular microlens field. There are also discontinuities in the critical curves when crossing the boundary of the rectangular region due to the complex logarithms and the boxcar function. However, the problem is not so easily rectifiable. We can recognize that the terms $-\kappa_\star B_{(c_1,c_2)}(z)$ are the convergence terms originating from the finite rectangular smooth mass sheet. The discontinuities arising from this $B_{(c_1,c_2)}(z)$ term can be removed by extending the mass sheet out to infinity once more. However, another set of discontinuities arises from branch cuts in the logarithms.

In addition, our ultimate goal is to arrive at some equation which we can find the roots of, in order to then parametrically trace out the critical curves. Unlike polynomials, quantifying the number of solutions for transcendental equations is more tricky, and we may not be able to ensure that we have found all the solutions desired. We therefore seek a different route, and take the sage advice of a friend to ``approximate ruthlessly''.

We can circumvent the problem by using a Taylor expansion of the complex logarithms in $\alpha_s(z)$. We want the expansion to be close to the true values within the rectangular region used for ray shooting, while it may be allowed to (and in truth will) be incorrect outside that region. We provide the details in Appendix \ref{app:a_smooth_approx}. In essence, $\alpha_s(z)$ is approximated by a polynomial. The expression for $\overline{\partial\alpha_z/\partial\overline{z}}$ is therefore also a polynomial of some degree, and the true number of roots $n_r$ to be found is thus somewhat greater than $2N_\star$.

This discussion serves only to highlight the idea that a finite field of microlenses introduces trouble in finding the critical curves near the boundary. The problem is alleviated for a circular microlens field, as symmetry allows us to expand the circular disk into an infinite mass sheet. The critical curves inside the field of microlenses are exact, while the critical curves outside the field are incorrect, but the calculated critical curves are continuous everywhere. A slightly more complicated approach must be taken when the microlens field is rectangular, but ultimately approximations can be made which alleviate the problem.

\section{Approximating the deflection angle of a rectangular mass sheet}
\label{app:a_smooth_approx}
By Taylor expanding around the origin, one has that 

\begin{equation}
   \log{(c-\overline{z})} = \log{c} - \sum_{j=1}^{\infty}\frac{1}{j}\Big(\frac{\overline{z}}{c}\Big)^{j}
\end{equation} which converges for $|z| < |c|$, i.e. within our rectangle (and in general, over a larger circular region) as desired. Similar expansions hold for $\overline{c}$, $-c$, and $-\overline{c}$. With this expansion and the extension of our finite rectangular mass sheet to an infinite one, we have

\begin{align}
    \begin{split}
    \alpha_s(z) =&\begin{aligned}[t]
        \frac{-\kappa_\star i}{\pi}\Big[&(c-\overline{z})\left(\log(c) - \sum_{j=1}^{\infty}\frac{1}{j}\left(\frac{\overline{z}}{c}\right)^{j}\right)\\
        &-(\overline{c}-\overline{z})\left(\log(\overline{c}) - \sum_{j=1}^{\infty}\frac{1}{j}\left(\frac{\overline{z}}{\overline{c}}\right)^{j}\right)\\
        &+(-c-\overline{z})\left(\log(-c) - \sum_{j=1}^{\infty}\frac{1}{j}\left(\frac{\overline{z}}{-c}\right)^{j}\right)\\
        &-(-\overline{c}-\overline{z})\left(\log(-\overline{c}) - \sum_{j=1}^{\infty}\frac{1}{j}\left(\frac{\overline{z}}{-\overline{c}}\right)^{j}\right)\Big]
    \end{aligned}\\
    &-\kappa_\star(c+\overline{c}+\overline{z} + z)
    \end{split}
\end{align} 

The Taylor expansions must of course be truncated at some point. The error $\epsilon$ between the approximation for $\alpha_s(z)$ truncated at some power $t$ and the actual value is \begin{equation}
    \begin{split}
    \epsilon =\frac{-\kappa_\star i}{\pi}\Big[&-(c-\overline{z})\sum_{j=t+1}^{\infty}\frac{1}{j}\left(\frac{\overline{z}}{c}\right)^j+ (\overline{c}-\overline{z})\sum_{j=t+1}^{\infty}\frac{1}{j}\left(\frac{\overline{z}}{\overline{c}}\right)^j\\
    &-(-c-\overline{z})\sum_{j=t+1}^{\infty}\frac{1}{j}\left(\frac{\overline{z}}{-c}\right)^j+(-\overline{c}-\overline{z})\sum_{j=t+1}^{\infty}\frac{1}{j}\left(\frac{\overline{z}}{-\overline{c}}\right)^j\Big]
    \end{split}
\end{equation} Using geometric series, the error can be bounded by \begin{equation}
    |\epsilon|\leq\frac{\kappa_\star}{\pi}\cdot \frac{4}{t+1}\left|\frac{z}{c}\right|^{t+1}\frac{|c| + |z|}{1-|z/c|} 
\end{equation} When shooting rays within the field of stars, $|z|\cdot s < |c|$ for some $s > 1$. We then have \begin{equation}
    |\epsilon|\leq\frac{\kappa_\star}{\pi}\cdot \frac{4}{t+1}\left|\frac{1}{s}\right|^{t+1}\cdot |c|\cdot \frac{s+1}{s-1}
\end{equation} from which, given some desired error $|\epsilon|$, $t$ can be found.

From the approximation for $\alpha(z)$ we can then find 

\begin{equation}
    \frac{\partial \alpha_s(z)}{\partial z} = -\kappa_\star
\end{equation} and, after lengthy algebra, 

\begin{equation}
    \begin{split}
    \frac{\partial \alpha_s(z)}{\partial \overline{z}} =&\begin{aligned}[t]
        \frac{-\kappa_\star i}{\pi}\Bigg[&\overline{z}^t\left(1+(-1)^t\right)\left(\frac{1}{c^t}-\frac{1}{\overline{c}^t}\right)\\
        & +\sum_{j=1}^{t}\frac{1}{j}\cdot\overline{z}^j\cdot\left(1 + (-1)^j\right)\left(\frac{1}{c^j} - \frac{1}{\overline{c}^j}\right)\Bigg]
    \end{aligned}\\
    &+\kappa_\star -\frac{4\kappa_\star \arctan{(c_2/c_1)}}{\pi}
    \end{split}
\end{equation}
Using $\overline{c}=c\cdot e^{-i\cdot2\cdot\arctan{(c_2/c_1)}}=c\cdot e^{-i\cdot2\cdot\Arg{c}}$, this can be further simplified to \begin{equation}
    \begin{split}
    \frac{\partial \alpha_s(z)}{\partial \overline{z}} =&\begin{aligned}[t]
        \frac{-\kappa_\star i}{\pi}\Bigg[&\left(\frac{\overline{z}}{c}\right)^t\left(1+(-1)^t\right)\left(1-e^{i\cdot 2\cdot\Arg{c}\cdot t}\right)\\
        &+\sum_{j=1}^{t}\frac{1}{j}\cdot\left(\frac{\overline{z}}{c}\right)^{j}\cdot\left(1 + (-1)^j\right)\left(1 -e^{i\cdot2\cdot\Arg{c}\cdot j}\right)\Bigg]
    \end{aligned}\\
    &+\kappa_\star -\frac{4\kappa_\star \Arg{c}}{\pi}
    \end{split}
\end{equation} which is more easily summed in code with Horner's method.

We can also evaluate

\begin{equation}
    \begin{split}
    \frac{\partial^2 \alpha_s(z)}{\partial \overline{z}^2} =&\begin{aligned}[t]
        \frac{-\kappa_\star i}{\pi \overline{z}}\Bigg[&t\left(\frac{\overline{z}}{c}\right)^t\left(1+(-1)^t\right)\left(1-e^{i\cdot 2\cdot\Arg{c}\cdot t}\right)\\
        &+\sum_{j=1}^{t}\left(\frac{\overline{z}}{c}\right)^{j}\cdot\left(1 + (-1)^j\right)\left(1 -e^{i\cdot2\cdot\Arg{c}\cdot j}\right)\Bigg]
    \end{aligned}
    \end{split}
\end{equation}

While these expressions approximating $\alpha_s(z)$ and its derivatives are complicated, they may be abstracted away in the code so that we need not dwell on them. The end result is that they add a polynomial of some degree to the expression for $F(z)$. 

One would assume, from the expressions above, that there are an additional $t$ roots. However, if $t$ is odd, the highest order term is 0 and so there are only $t-1$ additional roots. Since that somewhat simplifies the expressions, we enforce $t$ odd without loss of generality. 

If $(t-1)\cdot\Arg{c}$ is an integer multiple of $\pi$ though, then there are yet fewer roots. To avoid issues with numerical precision checking whether $(t-1)\cdot\Arg{c}$ is an exact integer multiple of $\pi$, we simply increase $t$ by 2 (to keep it odd) until $(t-1)\cdot\Arg{c}\mod\pi$ is sufficiently far from $\pi$ (by $\sim$10\%) as to not cause an issue. 

\section{Maximum error in \texorpdfstring{$1/\mu$}{1 / μ}}
\label{app:max_mu_error}

We provide an expression for the maximum error in $1/\mu$ based on the value of $F(z)$ at a particular point. We have 

\begin{equation}
    \begin{split}
    1/\mu =& \left(1-\kappa - \frac{\partial\alpha_s}{\partial z}\right)^2 \\
    &- \left(\gamma+\theta_\star^{2}\sum_{i=1}^{N_\star}\frac{m_{i}}{(z-z_i)^2}-\overline{\frac{\partial\alpha_s}{\partial \overline{z}}}\right)\left(\gamma+\theta_\star^{2}\sum_{i=1}^{N_\star}\frac{m_{i}}{(\overline{z}-\overline{z}_i)^2}-\frac{\partial\alpha_s}{\partial \overline{z}}\right)\\
    =&\left(1-\kappa - \frac{\partial\alpha_s}{\partial z}\right)^2\\
    &- \left(F(z)+\left(1-\kappa - \frac{\partial\alpha_s}{\partial z}\right)e^{-i\phi}\right)\left(\overline{F(z)}+\left(1-\kappa - \frac{\partial\alpha_s}{\partial z}\right)e^{i\phi}\right)\\
    =& -\big|F(z)\big|^2 - F(z)\left(1-\kappa - \frac{\partial\alpha_s}{\partial z}\right)e^{i\phi}-\overline{F(z)}\left(1-\kappa - \frac{\partial\alpha_s}{\partial z}\right)e^{-i\phi}\\
    =&-\big|F(z)\big|^2 - 2\big|F(z)\big|\left(1-\kappa - \frac{\partial\alpha_s}{\partial z}\right)\cos{(\phi+f)}
\end{split}
\end{equation} where we have used \begin{equation}
    F(z) = \big|F(z)\big|e^{if}
\end{equation} for some argument $f$. The maximum value of the error from the critical curve, where $1/\mu=0$, is then one of \begin{equation}
    \big|1/\mu\big| = \left|\left|F(z)\right|^2 \pm 2\left|F(z)\right|\left(1-\kappa - \frac{\partial\alpha_s}{\partial z}\right)\right|
\end{equation}


\bsp	
\label{lastpage}

\end{document}